\input epsf

\documentstyle[prl,floats,aps]{revtex}

\tighten

\begin{document}

\draft

\twocolumn[\hsize\textwidth\columnwidth\hsize\csname@twocolumnfalse\endcsname

\title{Three-dimensional simulations of distorted Black Holes.  I.
Comparison with axisymmetric results.}

\author{Karen Camarda${}^{(1)}$ and Edward Seidel${}^{(2,3,4)}$}
\address{
${}^{(1)}$
Department of Astronomy and Astrophysics and Center for
Gravitational
Physics and Geometry, \\
Pennsylvania State University, University Park, PA
16802
}
\address{
${}^{(2)}$ Max-Planck-Institut f{\"u}r
Gravitationsphysik,
Schlaatzweg 1, 14473 Potsdam,
Germany
}
\address{
${}^{(3)}$ National Center for Supercomputing
Applications,
Beckman Institute, 405 N. Mathews Ave., Urbana, IL
61801
}
\address{
${}^{(4)}$ Departments of
Astronomy and Physics,
University of Illinois, Urbana, IL 61801
}

\date{\today}
\maketitle

\begin{abstract}
We consider the numerical evolution of black hole initial data
sets, consisting of single black holes distorted by strong
gravitational waves, with a full 3D, nonlinear evolution code.  These
data sets mimic the late stages of coalescing black holes.  We compare
various aspects of the evolution of axisymmetric initial data sets,
obtained with this 3D code, to results obtained from a well
established axisymmetric code.  In both codes we examine and compare
the behavior of metric functions, apparent horizon properties, and
waveforms, and show that these dynamic black holes can be accurately
evolved in 3D. In particular we show that with present computational
resources and techniques, the process of excitation and ringdown of
the black hole can be evolved, and one can now extract accurately the
gravitational waves emitted from the 3D Cartesian metric functions,
even when they carry away only a small fraction ($<< 1\%$) of the rest
mass energy of the system.  Waveforms for both the $\ell=2$ and the
much more difficult $\ell=4$ and $\ell=6$ modes are computed and
compared with axisymmetric calculations.  In addition to exploring the
physics of distorted black hole data sets, and showing the extent to
which the waves can be accurately extracted, these results also
provide important testbeds for all fully nonlinear numerical codes
designed to evolve black hole spacetimes in 3D, whether they use
singularity avoiding slicings, apparent horizon boundary conditions,
or other evolution methods.

\end{abstract}
\vskip2pc]

\section{Introduction}

\label{sec:Introduction}

In a recent paper\cite{Camarda97b}, we showed that full 3D, Cartesian
coordinate simulations of (axisymmetric) distorted, dynamic black
holes can be performed, and through a comparison with results obtained
with an axisymmetric code, that the waveforms can be accurately
obtained.  Although with present resolutions the length of the
simulations is limited to about $t \approx 40 M_{ADM}$ (due to the use of
singularity avoiding time slicings, as detailed below), these results
represent a step forward towards obtaining useful waveforms from 3D
numerical calculations that will be needed for detecting and analyzing
signals buried in the data collected by gravitational wave
observatories.  The need for an increasingly complex sequence of such
simulations leading to realistic coalescence calculations is great, as
gravitational wave observatories may detect such waves in about five
years\cite{Abramovici92}.  As black hole collisions are presently
considered a most likely source of signals to be detected initially by
these observatories, it is crucial to have a detailed theoretical
understanding of the coalescence process that can only be achieved
through numerical simulation\cite{Flanagan97a,Flanagan97b}.  In
particular, it is most important to be able to simulate accurately the
excitation of the coalescing black holes, to follow the waves
generated in the process, and to extract gravitational waveforms
expected to be seen by detectors.

These are very difficult calculations, as one must simultaneously deal
with singularities inside the black holes, follow the highly nonlinear
regime in the coalescence process taking place near the horizons, and
also calculate the linear regime in the radiation zone where the waves
represent a very small perturbation on the background spacetime
metric.  The energy carried by these waves is typically found to be on
the order of $10^{-6}-10^{-2}M_{ADM}$.  At such low amplitude, both
the generation and propagation of these signals are susceptible to
small numerical errors inherent in numerical simulations.
Furthermore, one must numerically extract these waves from the metric
functions actually being evolved, which in 3D are usually the
Cartesian metric functions (e.g., $g_{xx}$, etc.).  As these metric
functions do not correspond directly to physical degrees of freedom,
but rather to the gauge in which they are evolved, the determination
of waves is provided by a complicated numerical procedure to isolate
gauge-invariant waveforms (described below), which can introduce
further numerical errors.

In axisymmetry, in the case of stellar collapse~\cite{Stark85} or
rotating collisionless matter~\cite{Abrahams94a} the evolution has
been followed through black hole formation and emitted $\ell=2$
waveforms have been studied.  In this paper we focus on vacuum black
holes, which exist in the initial data.  Such data sets are likely to
be used in the fully general study of 3D black hole coalescence.
Evolutions of vacuum black hole data sets have been carried out
successfully for distorted black holes with
rotation~\cite{Brandt94b,Brandt94c} and
without~\cite{Abrahams92a,Anninos93c}, and for equal mass colliding
black holes~\cite{Anninos93b,Anninos94b}, but with difficulty.  These
2D evolutions can be carried out to roughly $t=100M_{ADM}$, where $M_{ADM}$
is the
ADM mass of the spacetime, although beyond this point large gradients
related to singularity avoiding slicings usually cause the codes to
become very inaccurate and crash.  In full 3D, the simulations are
more difficult as the numerical problems are exacerbated
\cite{Anninos94c}.  We will show examples of this difficulty in
sections below.

These problems should be reduced through the use of apparent horizon
boundary conditions (AHBC), that allow one to avoid the singularities
not by pathological slicing conditions, but by excising the region
inside the horizon.  To date, impressive successes have been achieved
in 1D~\cite{Seidel92a,Anninos94e,Scheel94,Marsa96}, and also with
spherical black holes in 3D~\cite{Anninos94c,Daues96a,Cook97a}.  But
these techniques have not yet been applied to dynamic, nonspherical
black hole systems.  A completely different approach, using
characteristic evolution, has recently been developed very
successfully for 3D single black hole evolution, including distorted
and rotating black holes~\cite{Gomez98a}, allowing extremely long term
3D evolutions for the first time.  But for truly dynamic black holes
evolved in 3D using any techniques (Cauchy evolution, Characteristic
evolution, or AHBC), careful extraction of waveforms has been only
recently been investigated in detail in Ref.~\cite{Camarda97b}.  Those
results showed for the first time that the black hole ringdown
dynamics, and the accompanying emission of low amplitude waves could
be accurately computed in a full 3D simulation.  Such results should
be a useful testbed for techniques such as characteristic evolution
and AHBC that enable long term black hole evolution.

In this paper we follow up on the results of Ref.\cite{Camarda97b}
with a much more extensive and detailed examination of the evolution
of axisymmetric black holes with a fully 3D code.  These black hole
initial data sets correspond to Schwarzschild initial data with a
nonlinear gravitational wave superimposed, creating a distorted
Schwarzschild system that will oscillate and settle down to a
spherical hole with waves traveling out far from the hole.  We show
that with current techniques and computational resources available to
3D numerical relativity, distorted black holes can be evolved through
the initial relaxation and part of the final ringdown period.  We
compare the evolution of metric functions and horizons obtained with
an axisymmetric code written in spherical-polar coordinates using
maximal slicing and a particular shift condition, to full 3D
evolutions with a very general 3D code written in Cartesian
coordinates, capable of using different slicing and shift conditions.
We show that the gravitational waveforms can be followed and
accurately extracted from the numerical evolutions, even though they
represent a small perturbation on the background spacetime which is
also being evolved.  We also study the extraction of fully {\em
nonaxisymmetric} modes with the present code.  In principle, of
course, with axisymmetric initial data one should not excite
nonaxisymmetric modes, but the extent to which they may be excited due
to numerical effects in present codes needs to be investigated.  We
show below that numerical effects may excite such modes to some degree.
This will provide important information when evolving true 3D,
nonaxisymmetric black hole data sets, so that true signals can be
clearly distinguished from numerical artifacts.

The extension of this work to full 3D black hole initial data, for
which no testbeds exist at present, is in progress and will be
published in the next paper in this series.  In particular,
nonaxisymmetric modes, such as the $\ell=2,m=2$ mode expected to be
important in the ringing radiation for rotating black holes at late
times~\cite{Flanagan97a}, and therefore an important signal for
gravitational wave observations, can now be
studied\cite{Camarda97a,Allen97a,Allen98a}.

The plan of the present paper is as follows.  In section \ref{sec:ivp}
we briefly review the initial data construction.  In section
\ref{sec:evolution}, we describe both the 2D and the 3D numerical
codes used in the present simulations, and we study the numerical
evolution of several representative distorted black hole data sets,
discussing the behavior of metric functions, horizons, and waveforms
extracted.  In section \ref{sec:conclusions} we summarize and describe
future work in this series.

\section{Initial Data}

\label{sec:ivp}
In this section we describe the axisymmetric initial data describing
the distorted black holes to be evolved.

\subsection{Mathematical construction and numerical implementation}
There is by now a large body of literature on black hole initial data
for numerical relativity.  For multiple black holes, the early
axisymmetric data sets of Misner\cite{Misner60} and Brill and
Lindquist\cite{Brill63} were generalized to include spin and momentum
by York and coworkers\cite{Bowen80}, culminating in the definitive
work in 3D of Cook\cite{Cook90,Cook91,Cook93}.  An alternate and
simpler technique for computing multiple black hole initial data was
recently developed by Brandt and Br{\"u}gmann\cite{Brandt97a}.

In this section we briefly describe the mathematical construction of
initial data evolved in this paper, which constitute yet another
family of black hole data sets for numerical relativity.  These data
sets are single, time symmetric, distorted black holes of the type
studied in Refs.\cite{Bernstein94a,Bernstein93a} as axisymmetric
initial data, and their evolutions have been studied extensively in
Ref.\cite{Abrahams92a}.  These data sets have also been extended to
include rotation\cite{Brandt94b,Brandt94c,Brandt94a}, but we will not
consider rotating black holes here.  We will make extensive use of
these data sets in our studies presented in this paper.  They consist
of a single black hole that has been distorted by the presence of an
adjustable torus of nonlinear gravitational waves, of a form
originally considered by Brill\cite{Brill59}, which surround it.  This
is discussed in much more detail in Ref.  \cite{Brandt97a}, which
focuses only on the initial data, where we study a broad family of
full 3D distorted black holes, with and without rotation.  We restrict
ourselves here to nonrotating and axisymmetric cases, as studied
previously in axisymmetry by \cite{Bernstein94a}.  The amplitude and
shape of the torus of waves can be specified by hand, as described
below, and can create very highly distorted black holes.  For our
purposes, we consider them as convenient initial data that create a
distorted black hole that mimics the merger, just after coalescence,
of two black holes colliding in axisymmetry \cite{Anninos94b}.

Following\cite{Bernstein94a}, we write the 3--metric in the form
originally used by Brill~\cite{Brill59}:
\begin{equation}
\label{eq:metric}
d\ell^2 =
\tilde{\psi}^4 \left( e^{2q} \left( d\eta^2 + d\theta^2 \right) +
\sin^2\theta d\phi^2 \right),
\end{equation}
where $\eta$ is a radial coordinate related to the standard
Schwarzschild isotropic radius, which we refer to here as $r$, by
\begin{equation}
r = \frac{M}{2}
e^{\eta}.
\end{equation}
Then the $\eta$ coordinate is related to the Cartesian coordinates by
\begin{equation}
	\sqrt{x^{2}+y^{2}+z^{2}} = e^{\eta}.
\label{coords}
\end{equation}
(Here, and in what follows, we set the scale parameter $M$
in\cite{Bernstein94a} to be 2.)  In spherical symmetry, if the ``Brill
wave'' function $q$ vanishes, the Schwarzschild initial data set
results, as shown below, so for small values of the Brill wave one may
regard this system as a perturbed black hole.  For large waves, the
black holes can become extremely distorted.

Given a choice for the ``Brill wave'' function $q$, the Hamiltonian
constraint leads to an elliptic equation for the conformal factor
$\tilde{\psi}$.  (As these data sets are time symmetric, the momentum
constraints are trivially satisfied.  In Ref.  \cite{Brandt97a} we
consider the non-time symmetric case.)  The function $q$ represents
the gravitational wave surrounding the black hole, and is chosen to be
\begin{equation}
\label{eq:q2d}
q\left(\eta,\theta,\phi\right) = a \sin^n\theta \left(
e^{-\left(\frac{\eta+b}{w}\right)^2} +
e^{-\left(\frac{\eta-b}{w}\right)^2} \right) \left(1+c
\cos^2\phi\right).
\end{equation}
Thus, an initial data set is characterized by the parameters
$\left(a,b,w,n,c\right)$, where, roughly speaking, $a$ is the
amplitude of the Brill wave, $b$ is its radial location, $w$ its
width, and $c$ and the even integer $n$ control its angular structure.
Note that we have generalized the original axisymmetric construction
to full 3D by the addition of the parameter $c$, but in this paper we
restrict ourselves to $c=0$ for comparison with axisymmetric results.
A study of full 3D initial data and their evolutions will be published
elsewhere \cite{Camarda97a,Brandt97a}.

If the amplitude $a$ vanishes, the undistorted Schwarzschild solution
results, leading to
\begin{equation}
\tilde{\psi} = 2 \cosh \left( \frac{\eta}{2}
\right).
\end{equation}
We note that just as the Schwarzschild geometry has an isometry that
leaves the metric unchanged under the operation $\eta \rightarrow
-\eta$, our data sets also have this property, even in the presence of
the Brill wave.  The radial point $\eta=0$ is called the throat, as it
is the surface that connects the identical geometries of two universes
connected by an Einstein-Rosen bridge, or wormhole.  As discussed in
\cite{Anninos94c,Bernstein93b}, this isometry condition can also be
applied during the evolution and in Cartesian coordinates as well.  We
will make use of this condition in the work described in this paper.

As discussed in Ref.~\cite{Brandt97a}, we compute the initial data
sets on a 3D grid in spherical coordinates, because the inner and
outer boundaries are easier to handle.  The outer boundary is a Robin
condition, and the inner boundary is provided by the isometry
condition described above.  Once we have the conformal factor, we can
specify the 3-metric, the form of which is given in
Eq.~(\ref{eq:metric}).  The extrinsic curvature vanishes, as these are
time-symmetric initial data sets.  Because of the higher resolution of
the spherical grid in the inner regions, we found that computing
derivatives of the conformal factor on this grid and storing the
interpolated values on the Cartesian grid leads to more accurate
evolution.  Thus, we compute the first and second derivatives of the
conformal factor $\tilde{\psi}$ with respect to the spherical
coordinates $(\eta,\theta,\phi)$ on the spherical grid.

For convenience, these manipulations are carried out on the spherical
grid, but we perform evolutions in Cartesian coordinates.  We have
found it adequate to interpolate these quantities from the very fine
2D spherical grid onto the Cartesian grid using linear interpolation.
With this information, we can perform simple coordinate
transformations to convert these quantities to Cartesian coordinates
obtaining:
\begin{equation}
\label{eq:init3dmet}
\gamma_{ij} = \psi^4 \left(
\begin{array}{ccc}
\frac{y^2+x^2 e^{2q}}{x^2+y^2} & \frac{xy(e^{2q}-1)}{x^2+y^2} & 0 \\
\frac{xy(e^{2q}-1)}{x^2+y^2} & \frac{x^2+y^2 e^{2q}}{x^2+y^2} & 0 \\
0 & 0 & e^{2q}
\end{array}
\right),
\end{equation}
using the coordinate transformation implicit in Eq.~(\ref{coords}).
Finally, $\psi$ and its Cartesian derivatives, computed via the
coordinate transformation, are stored for use in computing
``conformal'' numerical derivatives, as described in
Ref.~\cite{Anninos94c}.

\subsection{Characteristics of sample initial data sets} In this
section, we single out three initial data sets whose evolution and
waveform emission are studied in detail in the following sections, and
we present some of their physical characteristics before going on to
evolutions.

In Table \ref{tab:datasetparams}, for each initial data set we give
the ADM mass $M_{ADM}$, the apparent horizon mass $M_{AH}$, and the
maximum radiation loss $MRL$, defined as in Ref.  \cite{Bernstein94a}
as
\begin{equation}
MRL = \frac{M_{ADM} - M_{AH}(t=0)}{M_{ADM}}.
\end{equation}
The $MRL$ is an upper limit on the amount of energy the spacetime can emit as
computed by considering the second law of black hole dynamics.  The
three cases studied range from slightly distorted, with an $MRL$ of
about $5\times 10^{-3} M_{ADM}$, to highly distorted, with an $MRL$ of
$7\times 10^{-2} M_{ADM}$.  Problems 1 and 2 have the same distortion
amplitude parameter, but different angular index $n$, leading to quite
different ratios of $\ell=4$ to $\ell=2$ content in the initial data.

\begin{table}
\begin{tabular}{cccccc}
Problem & $a$  & $n$ & $M_{ADM}/M$ & $M_{AH}/M$ & MRL \\ \hline
1       & 0.5  & 2   & 0.92        & 0.86       & $7.0\times 10^{-2}$ \\
2       & 0.5  & 4   & 0.97        & 0.92       & $6.0\times 10^{-2}$ \\
3       & 0.1  & 4   & 0.97        & 0.96       & $4.8\times 10^{-3}$ \\
\end{tabular}
\caption{We give the parameters for the three initial data sets from
whose evolution we will present results in this paper.  The complete
set of black hole initial data parameters are $(a,b,w,n,c)$.  The
amplitude $a$ and angular index $n$ are given, while for all cases the
Brill wave location and width parameters are $b=0$, $w=1$, and $c=0$
indicating axisymmetric initial data.  The quantity $MRL$ is the
maximum radiation loss possible during evolution for each data set, as
discussed in the text.  $M$ is the scale parameter described in the
text, taken to be 2 in this work.}
\label{tab:datasetparams}
\end{table}

\begin{figure}
\epsfxsize=200pt \epsfbox{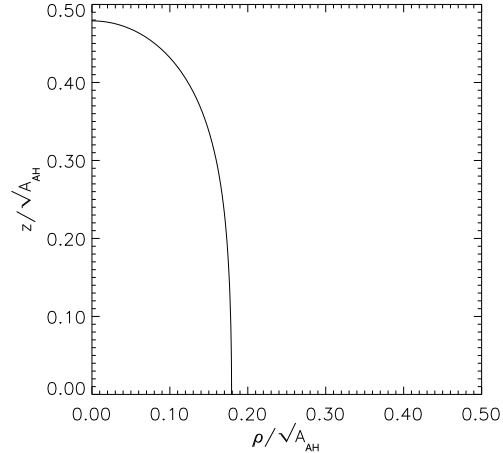}
\caption{We show the embedding
of the apparent horizon for Problem 1, $(a,b,w,n,c)=(0.5,0,1,2,0)$.
This black hole horizon is quite prolate, with a polar to equatorial
circumference ratio $C_{r}=2.7$}
\label{fig:run1.embd}
\end{figure}

\begin{figure}
\epsfxsize=200pt \epsfbox{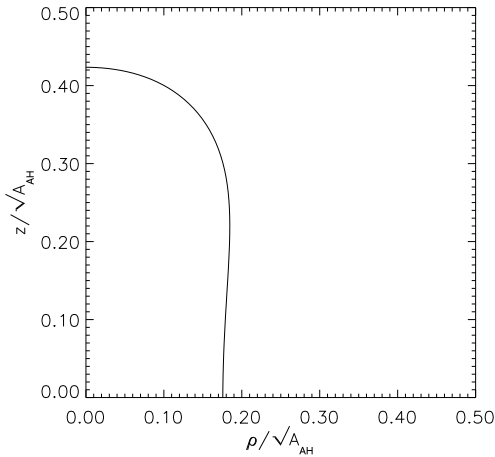}
\caption{We show the embedding
of the apparent horizon for Problem 2, $(a,b,w,n,c)=(0.5,0,1,4,0)$.
This horizon has a waist at the equator, with a polar to equatorial
circumference ratio $C_{r}=2.4$}
\label{fig:run2.embd}
\end{figure}

\begin{figure} \epsfxsize=200pt \epsfbox{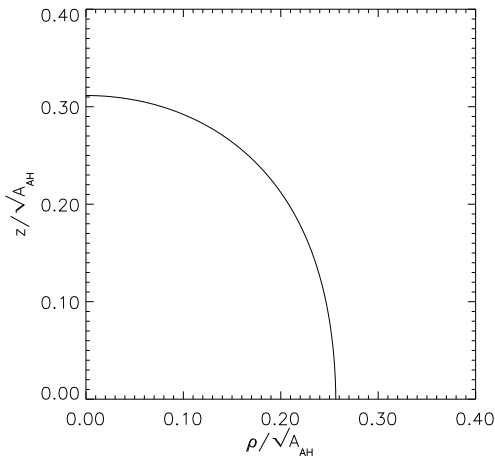}
\caption{We
show the embedding of the apparent horizon for Problem 3,
$(a,b,w,n,c)=(0.1,0,1,4,0)$.  This sightly prolate horizon is the
least distorted of the three Problems studied.}
\label{fig:run3.embd}
\end{figure}

To give a visual picture of how distorted the black holes in these
initial data sets are, we show the embedding of their apparent
horizons.  In Figs.~\ref{fig:run1.embd}, \ref{fig:run2.embd}, and
\ref{fig:run3.embd} we show the apparent horizon embeddings for
Problems 1, 2, and 3, respectively.  As shown previously in
Ref.~\cite{Bernstein94a}, black holes of this type with positive
distortion amplitude $a$ are have prolate horizons.  Problem 3 is
slightly perturbed, and nearly spherical, while Problems 1 and 2 are
noticeably prolate.  As in previous papers, we can characterize the
shape of axisymmetric black holes with a single parameter: the polar
to equatorial circumference ratio $C_{r}$.  The horizons for Problems
1 and 2 are quite prolate, with $C_{r}=2.7$ and $2.4$, respectively.
Problem 2, with its angular index $n=4$, actually has a ``waist'',
with negative Gaussian curvature in the region near the equator.
Axisymmetric (2D) evolutions of similar data sets were studied
extensively in Refs.~\cite{Abrahams92a,Anninos93c,Bernstein94a}, and
dynamics of the apparent and event horizons horizons of such data sets
were detailed in~\cite{Anninos93a,Anninos93d,Anninos94f,Anninos95c}.

These data sets will all go through oscillations, eventually settling
down to become spherical Schwarzschild black holes as energy is
radiated away.  Note that Problems 1-3 form a sequence from most to
least distorted, based on their horizon geometries, and also from
largest to least possible radiation output, measured by their $MRL$.
While Problem 3 is slightly distorted and a good candidate for study
via perturbation theory, Problems 1 and 2 are fairly strongly
distorted and are not well described by a first order perturbative
treatment\cite{Allen97a,Allen98a}.  In the next section will study the
full 3D evolution of these and other data sets in Cartesian
coordinates.

\section{Evolution of Axisymmetric Initial Data Sets}
\label{sec:evolution}
We now turn to the 3D evolution of the series of axisymmetric initial
data sets described above.  Comparisons of these results with those
from well established 2D codes will help us confirm various aspects of
the 3D results.

\subsection{Numerical codes}
\label{sec:code}
\subsubsection{2D code}
\label{sec:2dcode}

The code used to produce the 2D results in this paper is based on the
one described in~\cite{Bernstein93b}.  This code uses a logarithmic
radial coordinate $\eta$ related to the standard Schwarzschild
isotropic radius $r$ by $\eta=\ln\left(2r/M\right)$, where $M$ is a
scale parameter.  As all data sets evolved with this code have
equatorial-plane symmetry, the polar angle $\theta$ was restricted to
lie between $0$ and $\pi/2$ to decrease computational time.  All 2D
results presented were obtained using $200$ radial zones and a maximum
radius of $\eta_{max}=6$, resulting in a radial grid spacing of
$\Delta\eta=0.03$.  The number of angular zones, chosen to make
$\Delta\theta\approx\Delta\eta$ was approximately 50.

The use of the logarithmic radial coordinate has the advantage of
providing fine resolution near the throat of the black hole and also
near the peak that develops in the radial metric function, while also
allowing the outer boundary to be placed far from the hole.
Disadvantages of this coordinate are that (i) The throat region
remains extremely well resolved, even after the horizon has moved
significantly away from the hole.  Therefore much computational effort
is wasted well inside the horizon where the lapse is typically near
zero and the region is causally disconnected from the outside.  (ii)
The grid becomes very coarse outside the horizon in the radiation
zone, because equal spacing in the $\eta$ coordinate leads to larger
and larger spacing in the more physical $r$ coordinate.  Under these
conditions, waves may be reflected back toward the black hole as they
are scattered off of the coarse grid at larger radii, as discussed
in~\cite{Abrahams92a,Bernstein93b}.  In fact, the effective $\Delta r$
at the location of our wave extraction in the 2D code is approximately
$0.45M$, which is typically much larger than the resolution of the 3D
grid.

Except for the tests performed with geodesic slicing, the slicing used
in the 2D code was maximal slicing.  In order to stabilize the code
along the axis, the shift vector is chosen to maintain the metric
function $\gamma_{\eta\theta}=0$, thus keeping constant-coordinate
lines perpendicular ($\gamma_{\eta\phi}$ and $\gamma_{\theta\phi}$
also vanish).  These details are extensively discussed in
Refs.~\cite{Anninos93c,Bernstein93b}.

\subsubsection{3D code}
\label{sec:3dcode}
\paragraph{The code.}
We developed a 3D code to study black holes and gravitational waves in
Cartesian coordinates.  This code (known as the ``G'' code) was
applied to Schwarzschild black holes, where we showed that using
singularity avoiding time slicings, a spherical black hole could be
evolved accurately to $t=30-50M$, depending on the resolution,
location of the outer boundary, and the slicing
conditions\cite{Anninos94c}.  Beyond that time, the code generally
crashes due to the unbounded growth of metric functions generated by
singularity avoiding slicings.  However, the focus
of~\cite{Anninos94c} was on spherical black holes, so no studies were
made of black hole oscillations and the waves that would be generated
in the process.  It was shown that with spherical initial data some
nonspherical behavior could be introduced by the Cartesian mesh and
boundary conditions, the numerics of which could in principle generate
spurious gravitational waves.  This 3D code was then applied to the
collision of two axisymmetric black holes (Misner
data)~\cite{Anninos96c}, where we showed by comparison to 2D results
that one could accurately track the merging of the horizons, and that
the radiation emitted was qualitatively the same, but at that time the
waveforms were not studied extensively.  Building on the work
presented in this paper and in Refs.~\cite{Camarda97b,Camarda97a}, a
more detailed study of the Misner data in 3D, including the waveforms,
is in preparation for publication elsewhere.

The same code was simultaneously applied to the problem of pure
gravitational waves\cite{Anninos94d,Anninos96b}, where many systems
were studied, from pure linear quadrupole waves to nonlinear waves,
and their propagation on a Cartesian mesh was studied.  In that study
it was shown that waves can be accurately evolved, although certain
problems with gauge modes in the ``near linear'' regime that can
confuse the results were identified, along with strategies to deal
with them.  Finally, the code has also been used to study coordinate
conditions in 3D numerical relativity~\cite{Balakrishna96a}.

Based on this ``G'' code, we have developed a new version using the
same numerical techniques, having similar convergence properties, etc.,
but rewritten to take advantage of newer parallel computer
architectures, notably the SGI/Cray Origin 2000.  We use the shared memory
paradigm for parallelism in this code, avoiding the need to use a
message passing library.  With this model, we find both efficient
single processor use and reasonably good scaling beyond 64 processors
on the Origin 2000.  The numerical techniques are as described in
detail in~\cite{Anninos94c}.  We evolve the 3-metric and extrinsic
curvature components with an explicit, staggered leapfrog scheme.  For
the algebraic slicing used to produce the results in this paper,
described in the next section, the lapse function is evolved in this
scheme as well.

\paragraph{Gauge conditions}
The 3D code has been written in a general way to accept general shift
and slicing conditions.  Although various shift conditions have been
tested (e.g. minimal distortion shift as reported in Ref.
\cite{Balakrishna96a} or an apparent horizon shift condition reported
in Ref.~\cite{Anninos94c}), in this work we restrict ourselves to the
zero shift case.  The application of different shift conditions to
black hole spacetimes in 3D will be presented elsewhere.

We have experimented with a number of different slicing conditions.
In Refs.~\cite{Anninos94c,Camarda97a} we discussed the use of geodesic
slicing, maximal slicing and a host of algebraic slicing conditions
for single and two black hole spacetimes.  Geodesic slicing is useful
only for performing certain tests of a code on a black hole spacetime.
In brief, the slicing condition that worked the best for the analytic
Schwarzschild and Misner two black hole spacetimes was the evolution
of an initially maximal lapse, which vanishes on the throat of the
black hole, with the ``1+log'' algebraic slicing.  With this slicing
condition we were able to produce accurate and stable evolutions for a
period of time long enough to study the ringdown of excited black
holes, and it is much more computationally efficient than maximal
slicing.

However, for the numerically generated distorted black hole initial
data sets there is no appropriate analytic antisymmetric (i.e.
vanishing on the throat) initial lapse to use, so we compute an
antisymmetric maximal lapse numerically.  Because it is inconvenient
to compute and antisymmetric lapse across the black hole throat (which
is a coordinate sphere) in the Cartesian coordinates with our elliptic
solver, we compute the antisymmetric maximal lapse on the spherical
grid and interpolate that onto the Cartesian grid.  We have shown that
this procedure when applied to the Schwarzschild case produces the
same results as when the analytic Schwarzschild lapse is used,
although it is necessary to use at least quadratic interpolation.
This initial lapse is then evolved with the algebraic ``1+log'' lapse
as described in Ref.~\cite{Anninos94c}.  This is the slicing condition
used in all runs in this paper.

\paragraph{Computational domain and boundaries}
For the outer boundary, we presently hold the metric functions fixed
in time.  If the boundary is placed too close, this can have a serious
effect on the behavior of waveforms and metric functions, and hence
the largest possible computational domain is desirable.  For
evolutions of $30-50M$, typically we need to place the outer boundary
at least as far as $20-25M$.  The inner boundary (on the throat) is
provided by an isometry condition, as described above.

Although the 3D evolution code is written without making use of any
symmetry assumptions, the initial data we evolve in this paper have
both equatorial plane symmetry and axisymmetry.  Hence we save on the
memory and computation required by evolving only one octant of the
system.  As shown in~\cite{Anninos94c}, this has no effect on the
simulations except to reduce the computational requirements by a
factor of eight.  Even with such computational savings, these can be
extravagant calculations.  Smaller simulations presented in this paper
have resolutions of typically $100^{3}$, and can easily be computed in
a few hours on a 32 processor Origin 2000 at AEI. The highest
resolution results presented in this paper were computed on a 3D
Cartesian grid of $300^{3}$ numerical grid zones, which is about a
factor three larger than the largest production relativity
calculations of which we are aware (which were about $200^{3}$ zones).
With our new code, these take about 12 Gbytes of memory, and require
about a day on a 128 processor, early access Origin 2000
supercomputer at NCSA. These calculations were carried out in the
summer of 1997 and winter of 1998.

\subsection{Metric functions}
In this section we make direct comparison between the quantities
actually being evolved in both the 2D and 3D codes: metric functions.
This provides a direct test of the evolutions of the same initial data
sets in completely different codes, coordinate systems, and in the 3D
case, without enforcing symmetries required in the 2D code.

\subsubsection{Geodesic slicing}
We start by considering evolutions of axisymmetric initial data sets
with geodesic slicing ($\alpha=1$).  For Schwarzschild, it is well
known that in geodesic slicing a point initially on the throat will
fall into the singularity at a time $t=\pi M$, and hence does not
allow long term evolutions.  For distorted black holes such as these,
the results will differ somewhat from this analytic result, but the
basic picture is the same.  As shown in Ref.  \cite{Anninos94c}
geodesic slicing provides a powerful tests of the evolution equations.
For these distorted black holes, it allows us to compare evolved
metric functions with those obtained from a 2D code without the
complication of a lapse calculation.

\begin{figure}
\epsfxsize=200pt \epsfbox{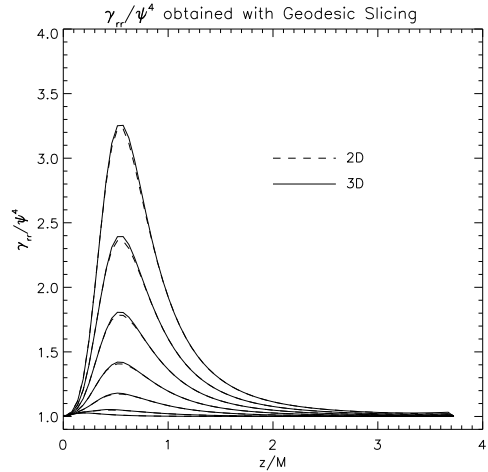} \caption{We show a comparison
of $\gamma_{rr}/\psi^4$ obtained with 2D and 3D codes using geodesic
slicing.  The data are shown every $0.163M_{ADM}$, to a maximum time
of $t=0.977M_{ADM}$.  The 3D data were obtained with $70^3$ grid
points with a resolution of $\Delta x=0.0543M_{ADM}$.  The data set
evolved was $(a,b,w,n)=(0.5,0,1,2)$ (Problem 1).}
\label{fig:grr2dcomp}
\end{figure}

\begin{figure}
\epsfxsize=200pt \epsfbox{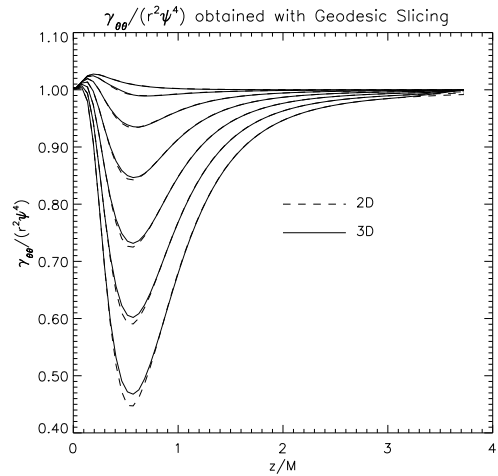} \caption{ As in Fig.
\protect{\ref{fig:grr2dcomp}}, except that we plot
$\gamma_{\theta\theta}/(r^2 \psi^4)$. The data are shown at the same
times as in Fig. \protect{\ref{fig:grr2dcomp}}. }
\label{fig:gtt2dcomp}
\end{figure}

In Fig.~\ref{fig:grr2dcomp} we show a plot of the conformal metric
function $\hat{\gamma}_{rr} = {\gamma}_{rr}/\psi^{4}$ obtained from 2D
and 3D codes along the $z$-axis for the distorted black hole initial
data set $(a,b,w,n,c)=(0.5,0,1,2,0)$.  Although the Cartesian metric
functions are the ones actually evolved, we reconstruct the spherical
metric functions numerically.  Furthermore, dividing out the time
independent conformal factor shows the dynamics more clearly.  It is
important to stress that the 2D data were obtained from an
axisymmetric code that uses the logarithmic radial coordinate $\eta$.
As discussed in section (\ref{sec:2dcode}), this has the effect of
providing very high resolution near the throat, unlike in the 3D case.
The 3D data were obtained using $70^3$ grid points and a resolution of
$\Delta x=0.0543M_{ADM}$.  In Fig.~\ref{fig:gtt2dcomp} we show a
comparison of $\hat{\gamma}_{\theta\theta}/r^2$ from the two codes for
the same runs.  In both plots we see that the metric functions line up
reasonably well.  The small differences can easily be explained by the
better resolution that is present in the 2D code, as discussed above.

The data in Figs.~\ref{fig:grr2dcomp} and~\ref{fig:gtt2dcomp} are only
shown to a time of $t=0.977M_{ADM}$.  The reason for this is not that
the throat observers have already fallen into the singularity at that
time.  In fact, the 3D code with that resolution can be run past
$t=1.95M_{ADM}$ without developing problems.  The problem is that the
2D code uses the standard spherical polar coordinate system, which has
a coordinate singularity on the axis.  Although there are no grid
points actually on the axis, the code becomes unstable for grid points
near the axis, causing the 2D code to crash very early.  This problem
is usually solved by using the gauge freedom of the Einstein equations
to use the shift vector to maintain the metric function
$\gamma_{r\theta}=0$, thus keeping constant-coordinate lines perpendicular.
However, the use of a non-zero shift results in a different 3-metric
than in the vanishing shift case.  Because this shift condition is not
convenient to implement in the 3D Cartesian code, we were forced to
compare with the zero-shift case, where the 2D code develops an axis
instability when the evolution becomes very dynamic there.

To summarize the results of this section, we have carefully compared
the evolutions of black hole initial data, in geodesic slicing, with
both 2D and 3D codes.  This provides a good test of the behavior of
the code in the strong field regions near the horizon and singularity.
In spite of the difficulties of evolution in 3D, in this case the 3D
evolution is more robust than the 2D case due to an axis instability.

\subsubsection{Singularity avoiding slicings}

We next consider evolutions with the ``1+log'' slicing condition.
The initial lapse is obtained as described above in
section~\ref{sec:3dcode}.  Here we only show that the metric functions
obtained are qualitatively the same in the two different codes.  We
cannot compare directly with data from 2D simulations, again because
of the different slicing and shift conditions used.

\begin{figure}
\epsfxsize=200pt \epsfbox{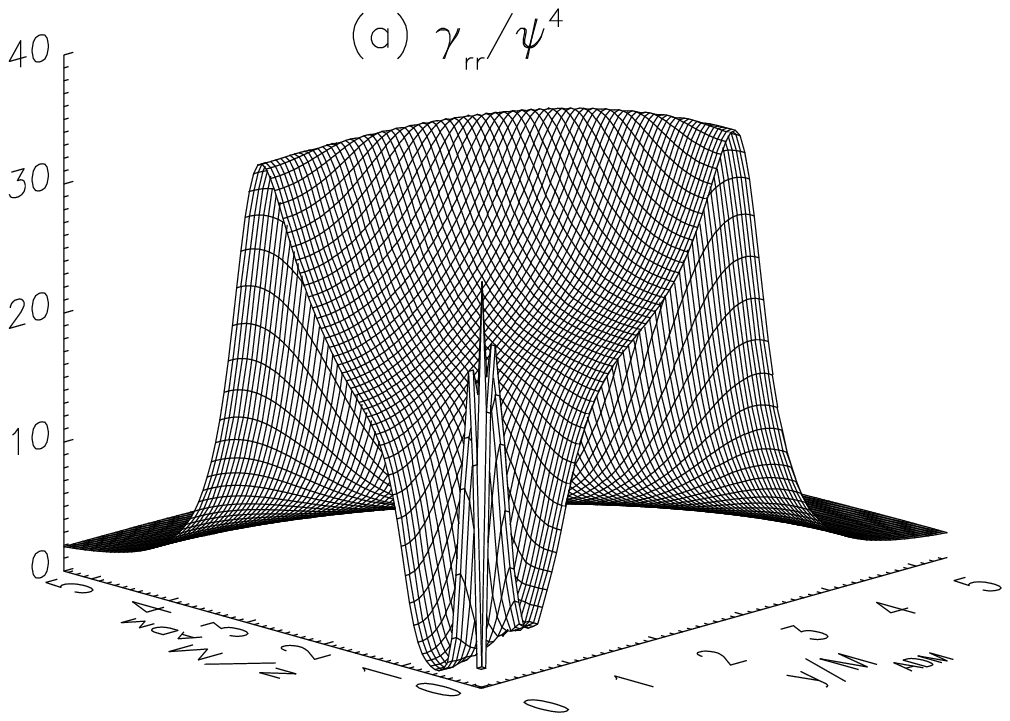} \epsfxsize=200pt
\epsfbox{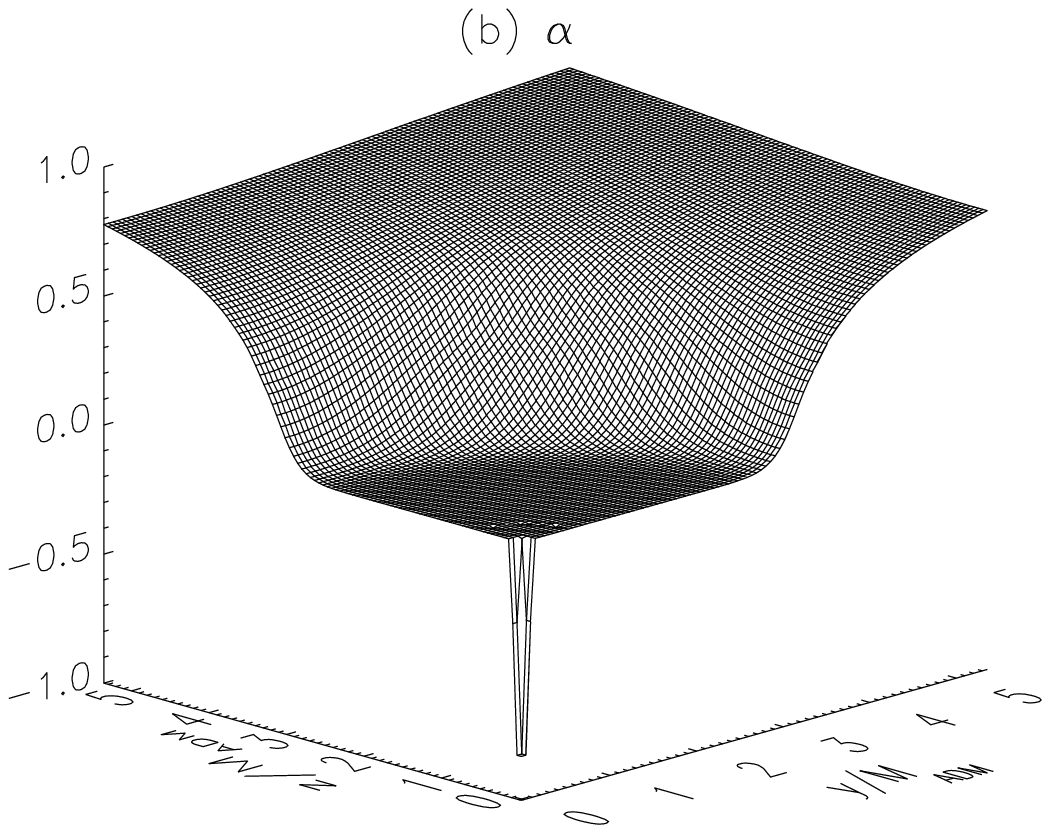} \caption{We show (a) the conformal metric
function $\hat{\gamma}_{rr}$ and (b) the lapse function $\alpha$ in
the $x=0$ plane for the evolution of the initial data set
$(a,b,w,n,c)=(0.5,0,1,2,0)$ (Problem 1).  The data shown are at a time
of $t=27.2M_{ADM}$.  The evolution was run on a $150^3$ grid, although
data from only the inner $100^3$ point are shown.  The resolution was
$\Delta x=0.0544M_{ADM}$.}
\label{fig:run1.met}
\end{figure}

As an example, we consider the evolution of the initial data set
$(a,b,w,n,c)=(0.5,0,1,2,0)$ (Problem 1).  This data set has an ADM
mass of $M_{ADM}=0.919M$.  The run was done with $150^3$ grid points
with a resolution of $\Delta x=0.0544M_{ADM}$.  In
Fig.~\ref{fig:run1.met}a, we show the metric function
$\hat{\gamma}_{rr}$ in the $x=0$ plane at the time $t=27.2M_{ADM}$.
We see the characteristic peak growing in this metric function, caused
by the differential acceleration of grid points towards the hole.  The
shape of the metric function is similar to that obtained with the 2D
code.  As in the Schwarzschild case, the peak growing near the origin
is a result of the application of the isometry condition.  In
Fig~\ref{fig:run1.met}b, we show the lapse function in the $x=0$ plane
at the same time.  We see that, as in the Schwarzschild and Misner
cases, the lapse which was initially antisymmetric at the throat, and
evolved with ``1+log'' slicing, has collapsed at and around the
throat.

The metric functions of the other black hole spacetimes studied in
this paper behave similarly, and the comparisons with 2D results
reveal the same qualitative agreement as expected.

\subsection{Horizons}
\label{sec:distbhah}

In this section we show a comparison of the location of the apparent
horizon (AH), computed in the 3D and 2D codes.  We computed the
location of the AH during an evolution of the initial data set
$(a,b,w,n,c)=(-0.5,0,1,2,0)$.  The calculation was performed with
$66^3$ grid points with a resolution of $\Delta x = 0.2 = 0.056M_{ADM}$.
For this run, maximal slicing was used.  The AH was found using the 3D
AH finder detailed in Ref.~\cite{Libson94b}.  This AH finder is
implemented in full 3D, based on an expansion of the 2D horizon
surface in symmetric trace free tensors.

\begin{figure}
\epsfxsize=200pt \epsfbox{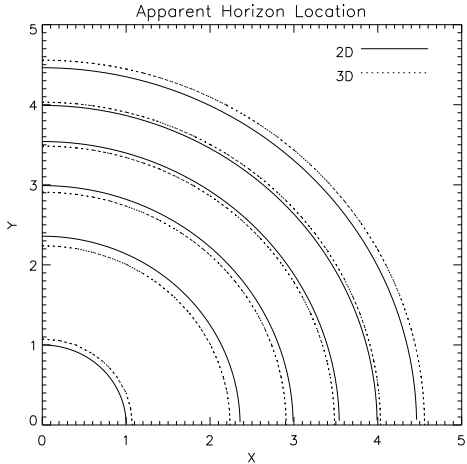} \caption{ Coordinate location of
the AH found in the 2D and 3D evolutions of the initial data set
$(a,b,w,n,c)=(-0.5,0,1,2,0)$.  The surfaces are shown starting at
$t=0$ with time intervals of $1.12M_{ADM}$.  Although we do not expect
identical results due to different kinematic conditions in the two
cases, the surfaces differ at most by little more than two grid cells.
The evolution is performed on a $66^3$ grid with $\Delta x = 0.2$.  }
\label{fig:dist_ev}
\end{figure}

Fig.~\ref{fig:dist_ev} shows the horizon shapes and locations in the
3D calculation every $1.15M_{ADM}$ in time, starting at $t=0$.  The
multipole order used in this calculation is $\ell=4$, with only
axisymmetric terms considered to reduce the computational time.  The
higher order expansion is needed here to describe the oblate shape of
the horizons.  Also shown in Fig.  \ref{fig:dist_ev} are the
corresponding surfaces found in our 2D axisymmetric code.  We note
that the slicing and shift conditions differ in the two cases, so we
do not expect the surfaces to coincide precisely.  In and around the
$x-y$ plane, the solutions match to within half a grid cell.  Greater
differences, however, are found along the $z$--axis, where the two
surfaces are displaced by a maximum of roughly two grid cells.  This
is attributed in part to a bigger $\Delta \alpha$ ($\sim 0.1$ ) near
the $z$--axis, which results from the imposed asymmetry in the lapse
function due to the nearness of the outer boundaries, where we enforce
the spherical Schwarzschild lapse as a boundary condition in the
maximal equation used in the 2D simulation.

\begin{figure}
\epsfxsize=200pt \epsfbox{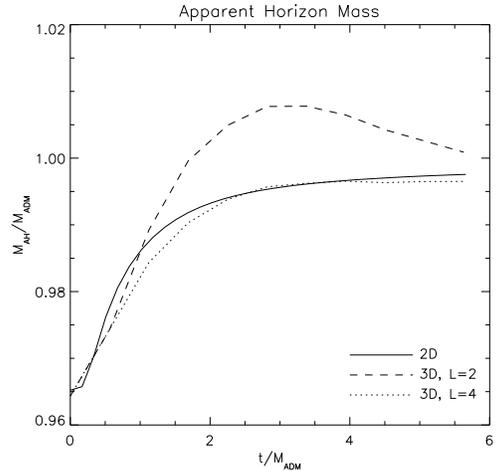}
\caption{ Comparison of the
apparent horizon masses computed from the 2D and 3D evolutions of the
initial data set $(a,b,w,n,c)=(-0.5,0,1,2,0)$.  The AH mass increases
initially as the Brill wave falls into the black hole.  By $t\sim 6M_{ADM}$,
the mass approaches $0.997M_{ADM}$ and the 2D and $\ell=4$, 3D results differ
by just 0.1\% at this time.  We also show the corresponding masses
computed from the surfaces found with an $\ell=2$ multipole expansion.
The low order expansion is clearly not adequate in resolving the
surface at early times when the horizon is most distorted.  }
\label{fig:dist_ev_ahm}
\end{figure}

A more geometrical comparison or test of the solver is the mass of the
surface found.  The horizon mass is defined as $M_{AH} =
\sqrt{A_{AH}/16\pi}$, where $A_{AH}$ is the area of the surface.  Fig.
\ref{fig:dist_ev_ahm} plots the AH mass as a function of time for the
2D and the $\ell=4$, 3D evolutions.  In both cases, the mass increases at
first, as the gravitational waves fall into the black hole, reaching
$M_{AH} \sim 0.997 M_{ADM}$ at $t\sim 6M_{ADM}$.  The masses in the two cases
differ by only 0.1\% at $t\sim 6M_{ADM}$.  For comparison we also plot
$M_{AH}$ for the surface found using a lower order $\ell=2$ multipole
expansion.  At early times, when the horizon is most distorted, the
$\ell=2$ expansion is clearly not adequate to resolve the horizon shape,
as evidenced by the AH mass which exceeds the ADM mass by about 1\%.
However, as the black hole settles down into a quasi--static state,
the surface becomes more spherical and the $\ell=2$ solution approaches
both the 2D and the $\ell=4$, 3D results, differing from the 2D result by
about 0.35\% at $t\sim 6M_{ADM}$.

\subsection{Radiation extraction}

\label{app:radex}
Although in black hole simulations we evolve directly the metric and
extrinsic curvature, for applications to gravitational wave astronomy
we are particularly interested in computing the waveforms emitted.
One measure of this radiation is the Zerilli function, $\psi$, which
is a gauge-invariant function that obeys the Zerilli wave
equation\cite{Zerilli70}.  The Zerilli function can be computed by
writing the metric as the sum of a spherically symmetric part and a
perturbation: $g_{\alpha\beta}=\stackrel{o}{g}_{\alpha\beta}+h_{\alpha
\beta}$, where the perturbation $h_{\alpha\beta}$ is expanded in
tensor spherical harmonics.  To compute the elements of
$h_{\alpha\beta}$ in a numerical simulation, one integrates the
numerically evolved metric components $g_{\alpha\beta}$ against
appropriate spherical harmonics over a coordinate 2--sphere
surrounding the black hole.  The resulting functions can then be
combined in a gauge-invariant way, following the prescription given by
Moncrief~\cite{Moncrief74}.  This procedure was originally developed by
Abrahams~\cite{Abrahams88}, and was applied to the same class of
distorted black hole initial data sets discussed here, but evolved in
2D spherical--polar coordinates and with a different gauge, as
discussed in~\cite{Abrahams92a}.

\subsubsection{3D numerical implementation}

We have developed numerical methods based on the same ideas to extract
the waves in a full 3D Cartesian setting.  The method used is
essentially that used in the axisymmetric case, except that the metric
functions and their spatial derivatives need to be interpolated onto a
two-dimensional surface, which we choose to have constant coordinate
radius.  The projections of the perturbed metric functions
$h_{\alpha\beta}$, and their radial derivatives, are then computed by
numerically performing two-dimensional surface integrals for each
$\ell-m$ mode desired.  Then, for each mode, the Zerilli function is
constructed from these projected metric functions, according to
Moncrief's gauge-invariant prescription.  This is a complicated but
straightforward procedure.

\paragraph{Constructing the Zerilli Function}
As mentioned above, we assume the general metric can be decomposed
into its spherical and non-spherical parts:
\begin{equation}
g_{\mu\nu} = \stackrel{o}{g}_{\mu\nu} +
h_{\mu\nu}.
\end{equation}
The spherical part $\stackrel{o}{g}_{\mu\nu}$ will of course be
Schwarzschild, but we will in general not know the mass of this
Schwarzschild background, or what coordinate system it will be in.
However, in general, we know it can be written
\begin{eqnarray}
\stackrel{o}{g}_{\mu\nu} =
\left(
\begin{array}{cccc}
-N^2 & 0 & 0 & 0 \\
0 & A^2 & 0 & 0 \\
0 & 0 &
R^2 & 0 \\
0 & 0 & 0 & R^2
\sin^2\theta
\end{array}
\right)
\end{eqnarray}
where the functions $N$, $A$, and $R$ are functions of our coordinate
radius and time.  Regge and Wheeler showed that $h_{\mu\nu}$ for
even-parity perturbations ({\it i.e.}, perturbations which do not
introduce angular momentum) can be written
\begin{eqnarray}
h_{tt} &=& -N^2 H_0^{(\ell m)} Y_{\ell m}
\\
h_{tr} &=& H_1^{(\ell m)} Y_{\ell m} \\
h_{t\theta} &=& h_0^{(\ell m)}
Y_{\ell m,\theta} \\
h_{t\phi} &=& h_0^{(\ell m)} Y_{\ell m,\phi} \\
h_{rr}
&=& A^2 H_2^{(\ell m)} Y_{\ell m} \\
h_{r\theta} &=& h_1^{(\ell m)} Y_{\ell
m,\theta} \\
h_{r\phi} &=& h_1^{(\ell m)} Y_{\ell m,\phi}
\\
h_{\theta\theta} &=& R^2 K^{(\ell m)} Y_{\ell m} + R^2 G^{(\ell m)}
Y_{\ell
m,\theta\theta}
\\
h_{\theta\phi} &=& R^2 G^{(\ell m)} \left(
Y_{\ell m,\theta\phi} -
\cot\theta Y_{\ell
    m,\phi} \right)
\\
h_{\phi\phi} &=& R^2 K^{(\ell m)} \sin^2\theta Y_{\ell m}
\nonumber \\
&&+ R^2
G^{(\ell
m)} \left( Y_{\ell
    m,\phi\phi} + \sin\theta \cos\theta Y_{\ell
m,\theta} \right)
\end{eqnarray}

The spherical part of the metric, given in the functions $N$, $A$, and
$R$, can be obtained by projecting the full metric against $Y_{00}$,
yielding the following expressions\footnote{We thank Gabrielle Allen
for pointing out the error in the expression for $R^2$ in
reference~\cite{Abrahams92a}.}:
\begin{equation}
N^2 = -\frac{1}{4\pi} \int g_{tt} d\Omega
\end{equation}
\begin{equation}
A^2 = \frac{1}{4\pi} \int g_{rr} d\Omega
\end{equation}
\begin{equation}
\label{eq:rschsq}
R^2 = \frac{1}{8\pi} \int \left( g_{\theta\theta} +
\frac{g_{\phi\phi}}{\sin^2\theta} \right) d\Omega
\end{equation}
Each $\ell m$-mode of $h_{\mu\nu}$ can then be obtained by projecting
the full metric against the appropriate $Y_{\ell m}$.  For example:
\begin{equation}
H_2^{(\ell m)} = \frac{1}{A^2} \int
g_{rr} Y_{\ell m} d\Omega
\end{equation}
Expressions for the other functions are provided in
Ref.~\cite{Allen97a}.  In practice, we do not extract $A$ to compute
$H_2^{(\ell m)}$, but rather we assume it to have the form
$1-\frac{2M}{R}$, where we take $M$ to be the ADM mass of the
spacetime.  For the axisymmetric, equatorial plane symmetric,
nonrotating initial data studied here, the only modes allowed are
even-parity, even-$\ell$ modes.

Moncrief showed that the Zerilli function is gauge invariant, and can
be constructed from the Regge-Wheeler variables as follows:
\begin{equation}
\psi^{(\ell m)} =
\sqrt{\frac{2(\ell-1)(\ell+2)}{\ell (\ell+1)}} \frac{4 R S^2
k^{(\ell m)}_2
+ \ell (\ell+1) R k^{(\ell
m)}_1}{\Lambda},
\end{equation}
where
\begin{eqnarray}
\Lambda &\equiv&
\ell (\ell+1) - 2 + \frac{6M}{R} \\
k^{(\ell m)}_1 &\equiv& K^{(\ell m)} +
S R G^{(\ell m)}_{,R} - 2 \frac{S}{R}
h^{(\ell m)}_1 \\
k^{(\ell m)}_2
&\equiv& \frac{H_2^{(\ell m)}}{2S} -
\frac{1}{2\sqrt{S}}
\frac{\partial}
{\partial R} \left( \frac{RK^{(\ell
m)}}{\sqrt{S}} \right) \\
S &\equiv& 1 - \frac{2M}{R}.
\end{eqnarray}

In order to compute the Regge-Wheeler perturbation functions $h_1$,
$H_2$, $G$, and $K$, one needs the spherical metric functions on some
2-sphere.  We get these by interpolating the Cartesian metric
functions onto a surface of constant coordinate radius, and computing
the spherical metric functions from these using the standard
transformation.  Second order interpolation is used.  In order to
compute the needed radial derivatives of these functions, we first
compute the derivative of the Cartesian metric functions with respect
to the coordinate radius on the Cartesian grid.  We interpolate these
quantities onto the 2-sphere.  Then, to get the derivatives with
respect to the Schwarzschild radius $R$, we use the derivative of
Eq.~(\ref{eq:rschsq}) with respect to the coordinate $r$, giving
\begin{equation}
\frac{\partial R}{\partial r} = \frac{1}{16\pi R}
\int \left(
g_{\theta\theta,r} + \frac{g_{\phi\phi,r}}{\sin^2\theta}
\right) d\Omega.
\end{equation}

The points on the 2-sphere to which we interpolate lie on lines of
constant $\theta$ and $\phi$, staggered across the coordinate axes.
Integrations over each of these variables is performed with the
second-order accurate Simpson's rule.  Currently, 300 points are used
for each angular coordinate.  The numerical interpolations involved in
this extraction procedure were also chosen to be second order
accurate.  Testing shows that both the interpolations and integrations
do converge to second order in the relevant grid spacing.

We are aware of two possible improvements on this scheme.  First, in
the $\theta$-integration, we could integrate over $\cos\theta$ instead
of $\theta$.  This could make these integrations more accurate, and
would allow us to use fewer points, making the code more efficient.
Second, we could interpolate onto a stereographic patch.  This would
distribute the points more evenly on the surface.

As in Ref.~\cite{Abrahams92a}, we choose to normalize the Zerilli
function so that the asymptotic energy flux in each mode is given by
\begin{equation}
\label{eq:energy}
\dot{E} = (1/32\pi) \dot{\psi}^2.
\end{equation}
This assumes that linear theory is adequate to treat each individual
mode.  However, it is possible that in some regimes linear theory is
{\em not} adequate, and hence the true energy calculation would
require additional terms.  The treatment of second order perturbation
theory of this sort, including a derivation of the energy formula
including higher order corrections, has been developed in
Ref.~\cite{Gleiser96a}.

Such considerations are actually relevant for the some of the initial
data sets described here.  For example, Problem 1 has angular
parameter $n=2$.  One can show~\cite{Brandt97a} that in an expansion
of the initial data for small amplitude $a$, one recovers the
Schwarzschild background with nonspherical perturbations, as assumed
above.  However, for $n=2$, the series expansion in the amplitude
parameter $a$ brings in the $\ell=2$ perturbations at linear order,
but $\ell=4$ does not appear until second order in $a$.  Therefore, in
order to be fully self-consistent the evolution equations for the
$\ell=4$ modes should actually be treated to second order, which
implies that there should be nonlinear source terms coming from other
linear modes (in this case, $\ell=2$) in a consistent perturbation
expansion, as discussed in Ref.~\cite{Gleiser96a}.  Basically, in such
cases the $\ell=4$ mode is so small, that nonlinear contributions from
other modes are not to be neglected.  This need not concern us,
because the evolutions considered in this paper are carried out fully
nonlinearly, but it does mean that the interpretation, and in
particular the calculation of the energy carried by a mode, must be
carefully considered.  Perturbative calculations, with comparisons to
full nonlinear simulations, are in progress and will be reported
elsewhere~\cite{Allen97a,Allen98a}.

In the results discussed below, in cases where second order
corrections should be considered in principle, we will continue to use
the linear energy formula as a non-rigorous indication of the strength
of a signal.  The cases in question concern the $\ell > n$ modes
(e.g., $\ell=6$ in all cases and $\ell=4$ for the $n=2$ simulations,
where $n$ is the angular index of the Brill wave.)

To summarize this section, we have developed a general technique to
extract the 3D waves propagating on a black hole background.  There
are cases where linear theory must be used with caution.  While
previously only axisymmetric simulations have been studied, we can now
study all non-trivial wave modes, including those with $m\ne 0$.

\subsubsection{Waveforms from 3D simulations of distorted black holes}

In this section, we examine the waveforms for the three test cases
described in detail above.  This is a detailed followup to
Ref.~\cite{Camarda97b}, so we present $\ell=2$ and $\ell=4$ waveforms
for Problem 1 as before.  We also consider two additional cases with
different amplitudes and angular index $n$ to show the fairly generic
nature of these results.  Finally, for the first time, we show that it
is now possible in some cases to extract accurately even the $\ell=6$
even-parity waveforms, both in the 2D and 3D cases, if the amplitude
is not too small.

\begin{figure}
\epsfxsize=200pt \epsfbox{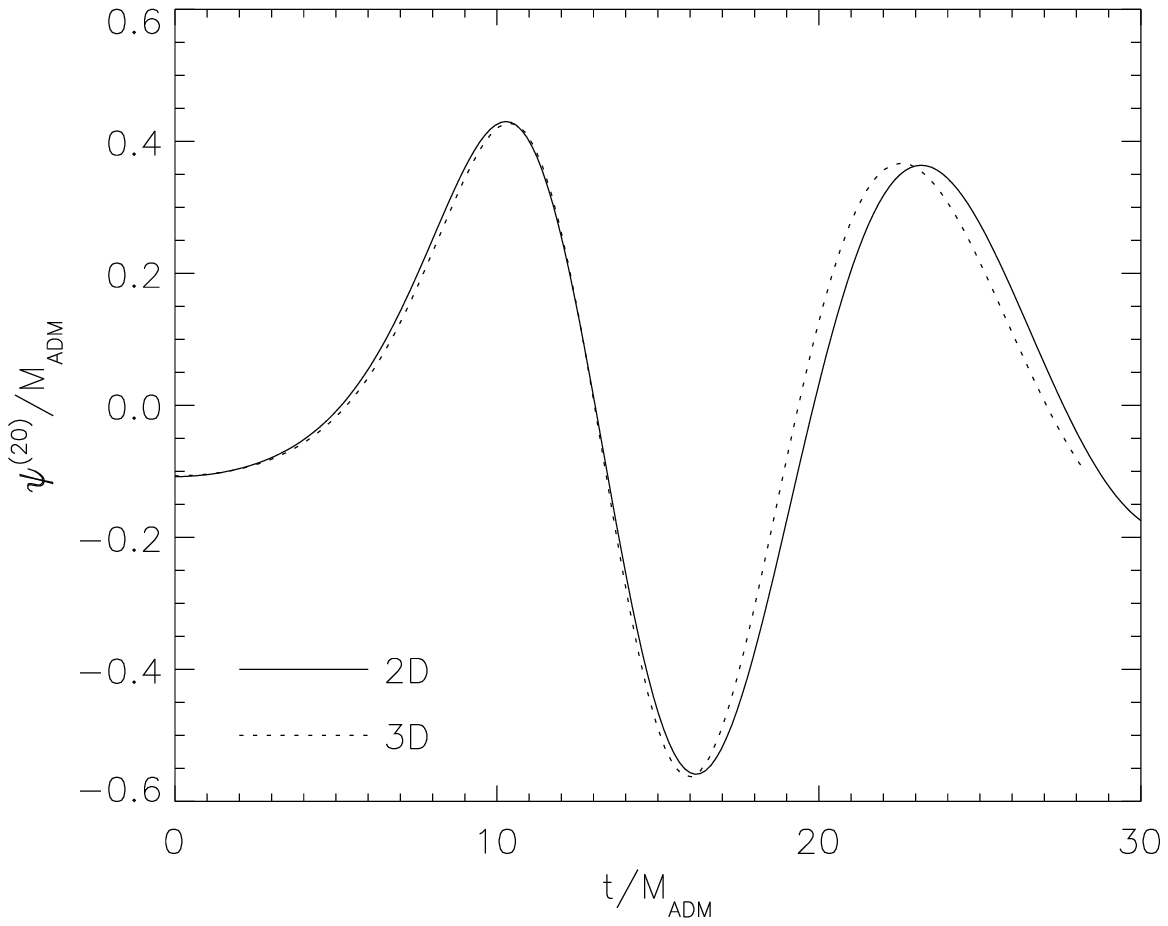}
\epsfxsize=200pt \epsfbox{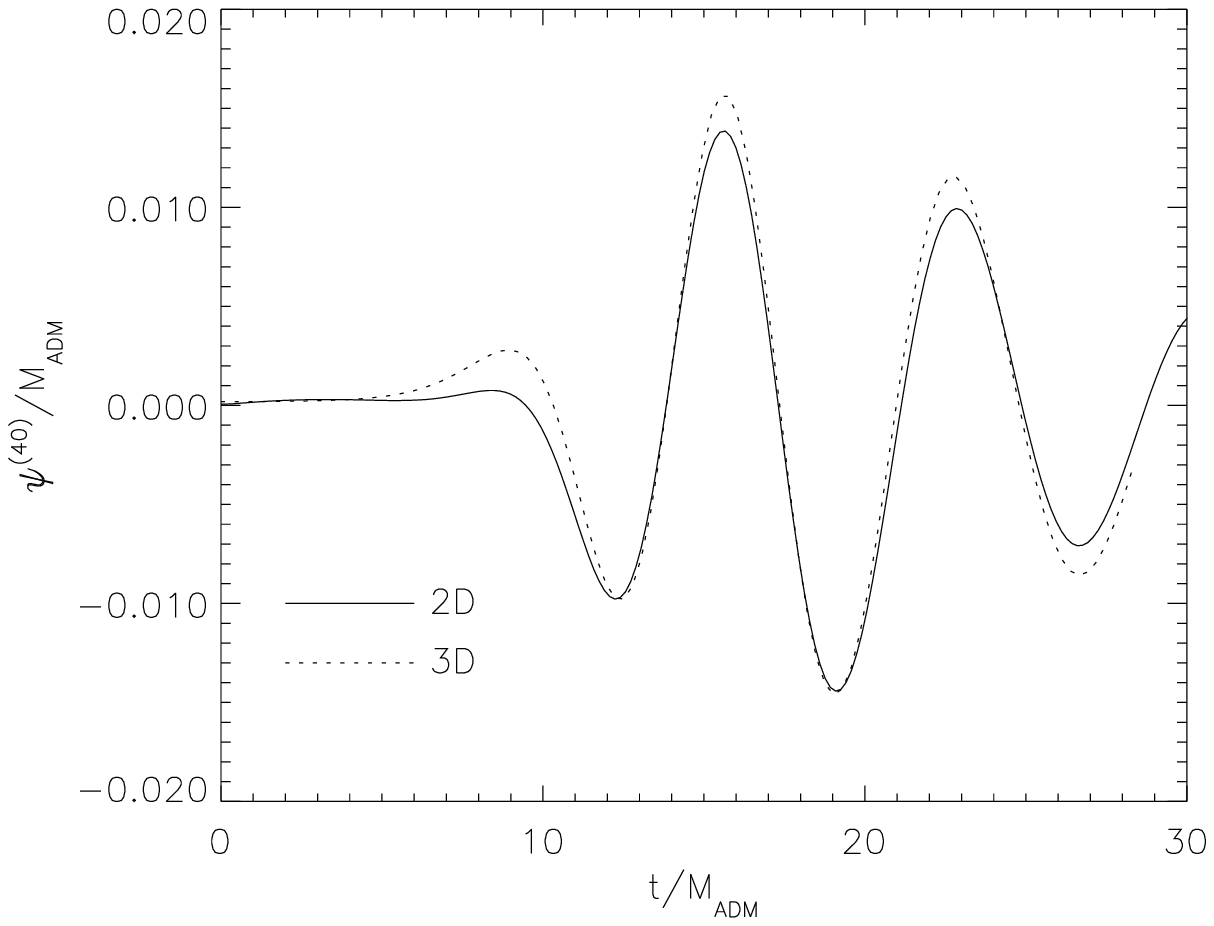}
\caption{We show the (a)
$\ell=2$, and (b) $\ell=4$ Zerilli functions vs.  time,
extracted during 2D and 3D evolutions of the data set
$(a,b,w,n,c)=(0.5,0,1,2,0)$ (Problem 1).  The functions were extracted
at a radius of $8.19M_{ADM}$.  The 2D data were obtained with
$202\times 54$ grid points, giving a resolution of
$\Delta\eta=\Delta\theta=0.03$.  The 3D data were obtained using
$300^3$ grid points and a resolution of $\Delta x=0.109M_{ADM}$.}
\label{fig:run1.zercomp}
\end{figure}

\paragraph{Problem 1.}
We extracted the $\ell=2$ and $\ell=4$ Zerilli functions during an
evolution of the distorted black hole initial data set $(a,b,w,n,c) =
(0.5,0,1,2,0)$, using the extraction method described above.  In
Fig.~\ref{fig:run1.zercomp}a we show the $\ell=2$ Zerilli function
extracted at a radius $r=8.16M_{ADM}$ as a function of time.
Superimposed on this plot is the same function computed during the
evolution of the same initial data set with a 2D code, based on the
one described in detail in \cite{Abrahams92a,Bernstein93b}.  The
agreement of the two plots over the first peak is a strong affirmation
of the 3D evolution code and extraction routine.  It is important to
note that the 2D results were computed with a different slicing
(maximal), different coordinate system, and a {\em different spatial
gauge}.  Yet the physical results obtained by these two different
numerical codes, as measured by the waveforms, are remarkably similar
(as one would hope).  This is a principal result of this paper.  A
full evolution with the 2D code to $t=100M_{ADM}$, by which time the
hole has settled down to Schwarzschild, shows that the energy emitted
in this mode at that time is about $4\times 10^{-3}M_{ADM}$.
Therefore, although this was a highly distorted black hole, less than
5\% of the $MRL$ is actually radiated away.  (Other modes are much
smaller and do not contribute significantly to the total energy
radiated.)

In Fig.~\ref{fig:run1.zercomp}b we show the $\ell=4$ Zerilli function
extracted at the same radius, computed during evolutions with 2D and
3D codes.  This waveform is more difficult to extract, because it has
a higher frequency in both its angular and radial dependence, and it
has a much lower amplitude: using the linear theory formula discussed
above (Eq.~(\ref{eq:energy})) the energy emitted in this mode is three
orders of magnitude smaller than the energy emitted in the $\ell=2$
mode, {\em i.e.}, $10^{-6}M_{ADM}$, yet it can still be accurately
evolved and extracted.

Small differences between the 2D and 3D results can be seen.
Resolution studies of the 3D results indicate that the differences are
not completely due to resolution of the 3D evolution code.  The small
differences in phase can be understood as a result of the different
shift and slicings being used in the two simulations.  The radiation
is extracted at a constant {\em coordinate} location, and the
coordinates fall towards the black hole at different rates with
different slicings and shifts.  By measuring the physical radial
position of the wave extraction in these simulations, we determined
that the difference between the 2D and 3D phases at late time is
consistent with the slightly different extraction locations in the two
cases.  The additional differences in the $\ell=4$ waveforms could be
related to slight differences in the initial data, which were
generated in independent ways, or even differences in gauge (the
waveforms are gauge-invariant, meaning they are unaffected only at
first order under gauge transformations).  As $\ell=4$ has a much
smaller amplitude than $\ell=2$, it will be more sensitive to such
details.

Since this data set is axisymmetric, the non-axisymmetric Zerilli
functions of a perfect evolution should vanish.  However, we
expect that numerical errors, especially errors due to the distinctly
non-axisymmetric outer boundary, will result in finite values for
these Zerilli functions.  Knowing the size of the error produced will
help us gauge the accuracy of our results when we move to full 3D,
where we expect real non-axisymmetric signals.

\begin{figure}
\epsfxsize=200pt \epsfbox{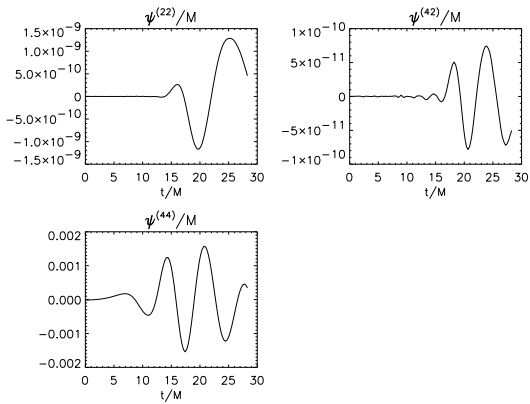}
\caption{We show the
non-axisymmetric Zerilli functions as a function of time extracted
during an evolution of the axisymmetric data set $(a,b,w,n,c)=
(0.5,0,1,2,0)$ (Problem 1).  The extraction was done at a radius of
$r=8.16M_{ADM}$.  The calculation was performed with $300^3$ grid
points and a resolution of $\Delta x=0.109M_{ADM}$.}
\label{fig:run1.nonaxi}
\end{figure}

In Fig.~\ref{fig:run1.nonaxi}, we show non-axisymmetric Zerilli functions
as a function of time, extracted during an evolution of the same data
set as above, at the same radius.  We see that the $\ell=2$, $m=2$ and
$\ell=4$, $m=2$ waveforms remain very small throughout the evolution.
The $\ell=4$, $m=4$ waveform, while an order of magnitude smaller than
the $\ell=4$, $m=0$ waveform, is still much larger than the other
non-axisymmetric waveforms.  Preliminary tests computing these
quantities for Schwarzschild initial data, for which all Zerilli
functions should vanish, indicate that the $\ell=4$, $m=4$ waveform is
more sensitive to errors in the interpolation of metric functions onto
the spherical extraction surface than the other Zerilli functions are.
The reason for this sensitivity and how to correct for it are under
investigation.

Resolution studies show that the waveforms are converging at roughly
second order in the grid spacing.  The axisymmetric $\ell=2,4$ modes
have converged at the resolution shown, while the $\ell=m=4$ mode is
converging towards zero, although at the highest resolution we have
computed it is still relatively large.

\paragraph{Problem 2}

\begin{figure}
\epsfxsize=200pt \epsfbox{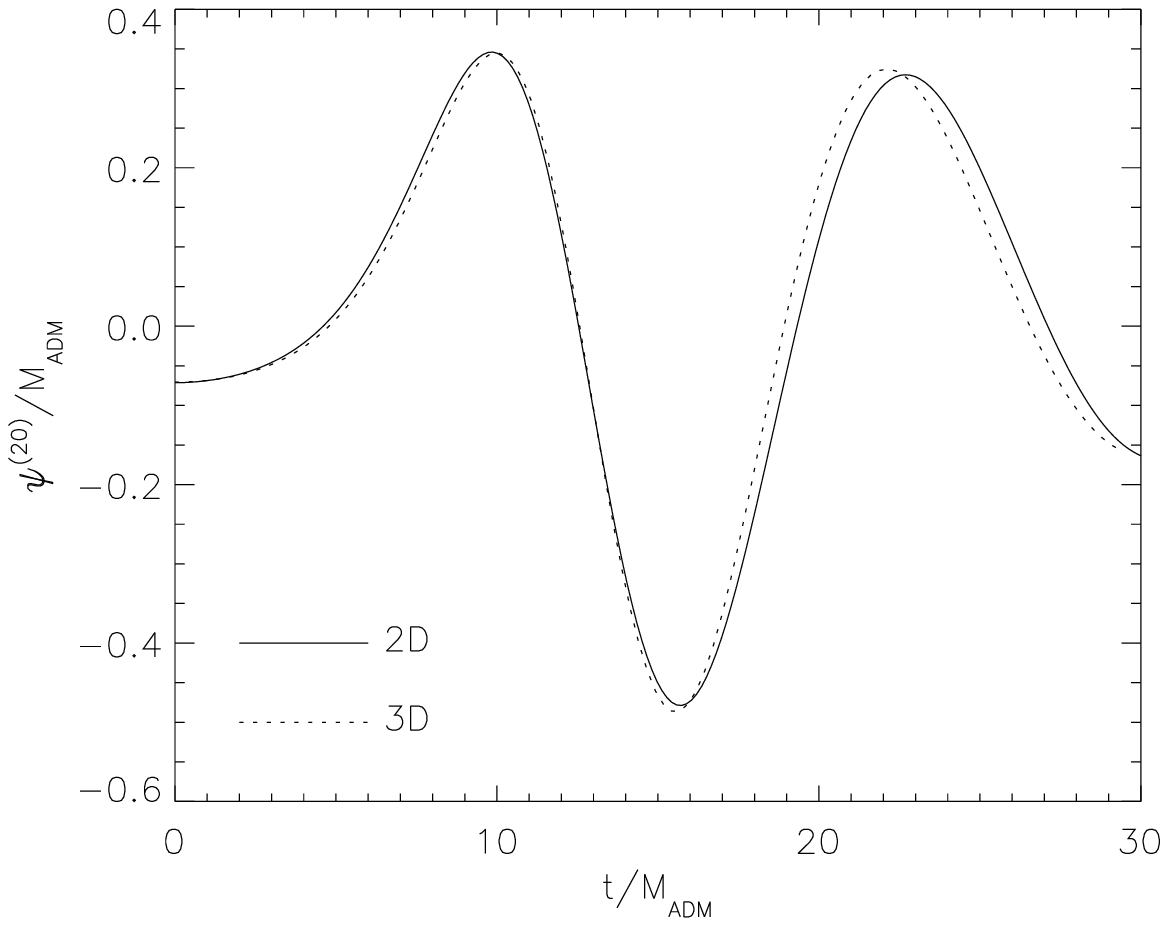}
\epsfxsize=200pt \epsfbox{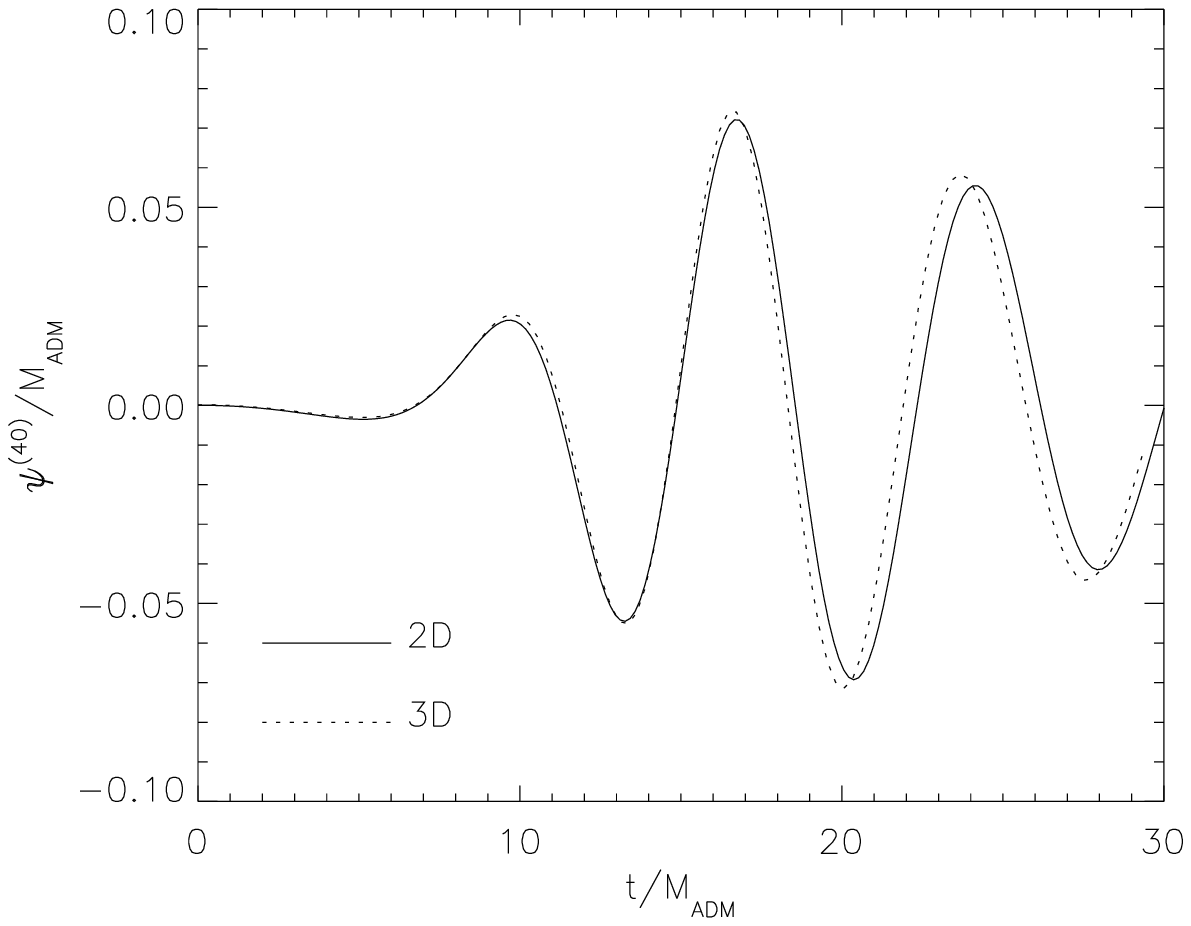}
\epsfxsize=200pt \epsfbox{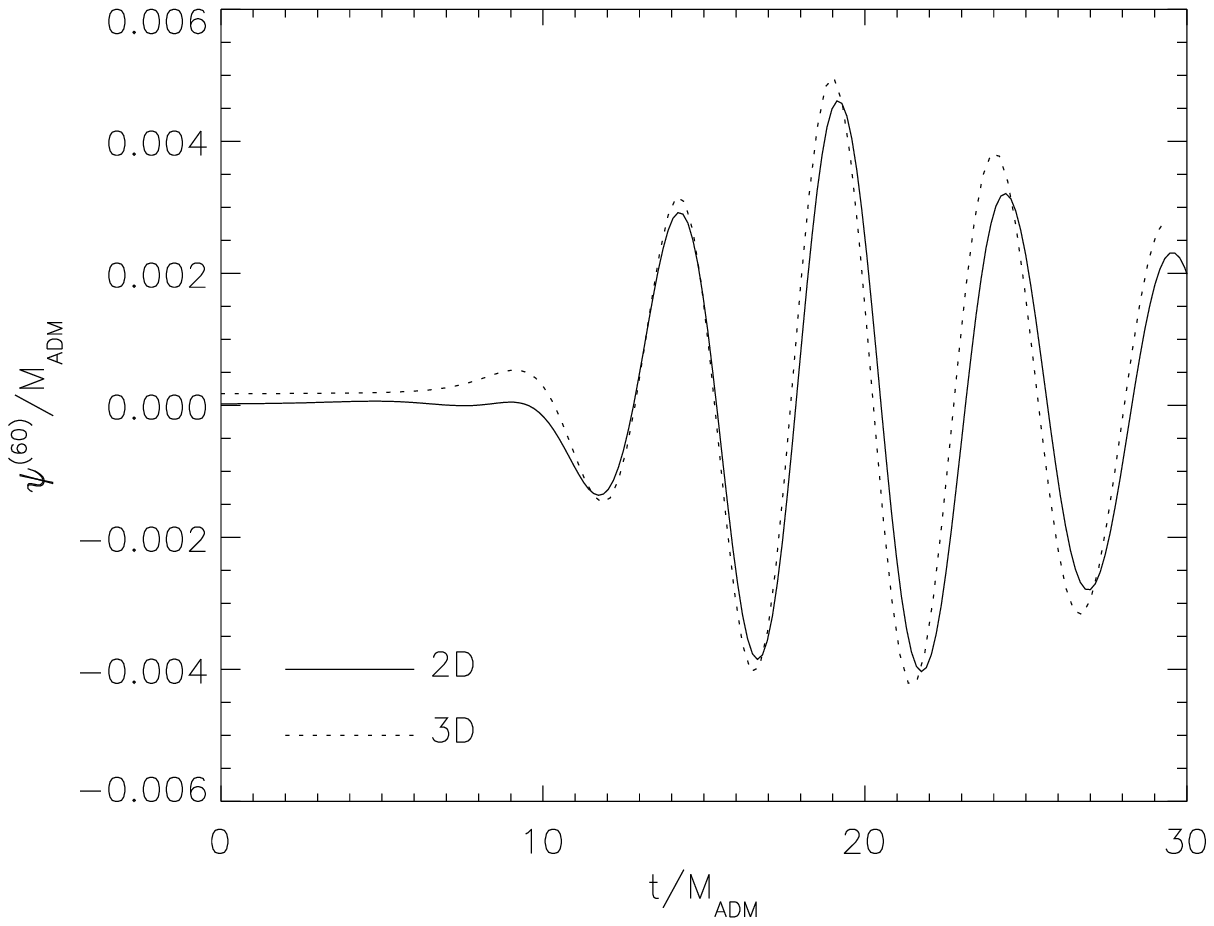}
\caption{We show the (a) $\ell=2$, (b) $\ell=4$, and (c) $\ell=6$
Zerilli functions vs.  time, extracted during 2D and 3D evolutions of
the data set $(a,b,w,n,c)=(0.5,0,1,4,0)$ (Problem 2).  The functions
were extracted at a radius of $7.72M_{ADM}$.  The 2D data were
obtained with $202\times 54$ grid points, giving a resolution of
$\Delta\eta=\Delta\theta=0.03$.  The 3D data were obtained using
$300^3$ grid points and a resolution of $\Delta x=0.103M_{ADM}$.}
\label{fig:run2.zercomp}
\end{figure}

We now turn to Problem 2, which is similar to Problem 1, but with
angular index $n=4$.  In this case the initial data contain a much larger
$\ell=4$ and $\ell=6$ mode content.  In this example, we extract the
$\ell=2,4,6$ modes during the 3D evolution, and compare with the 2D
axisymmetric evolution.  Fig.~\ref{fig:run2.zercomp} shows the
comparisons between the waveforms, showing that for this highly
distorted black hole, even the $\ell=6$ waveform computed in 3D agrees
very well with the axisymmetric simulation.  As in the previous case,
we emphasize that these two calculations are done in different
geometries, different slicing conditions, and different spatial
gauges.

For this simulation, the total radiated energy carried in the $\ell=2$
mode is $2\times 10^{-3} M_{ADM}$, or about 3\% of the MRL. The
$\ell=4$ mode carries $1\times 10^{-4} M_{ADM}$ in energy, much larger
than in the previous case.  In this case the $\ell=4$ mode is a first
order perturbation in the Brill wave amplitude, and the linear energy
calculation applies.

\paragraph{Problem 3}

\begin{figure}
\epsfxsize=200pt \epsfbox{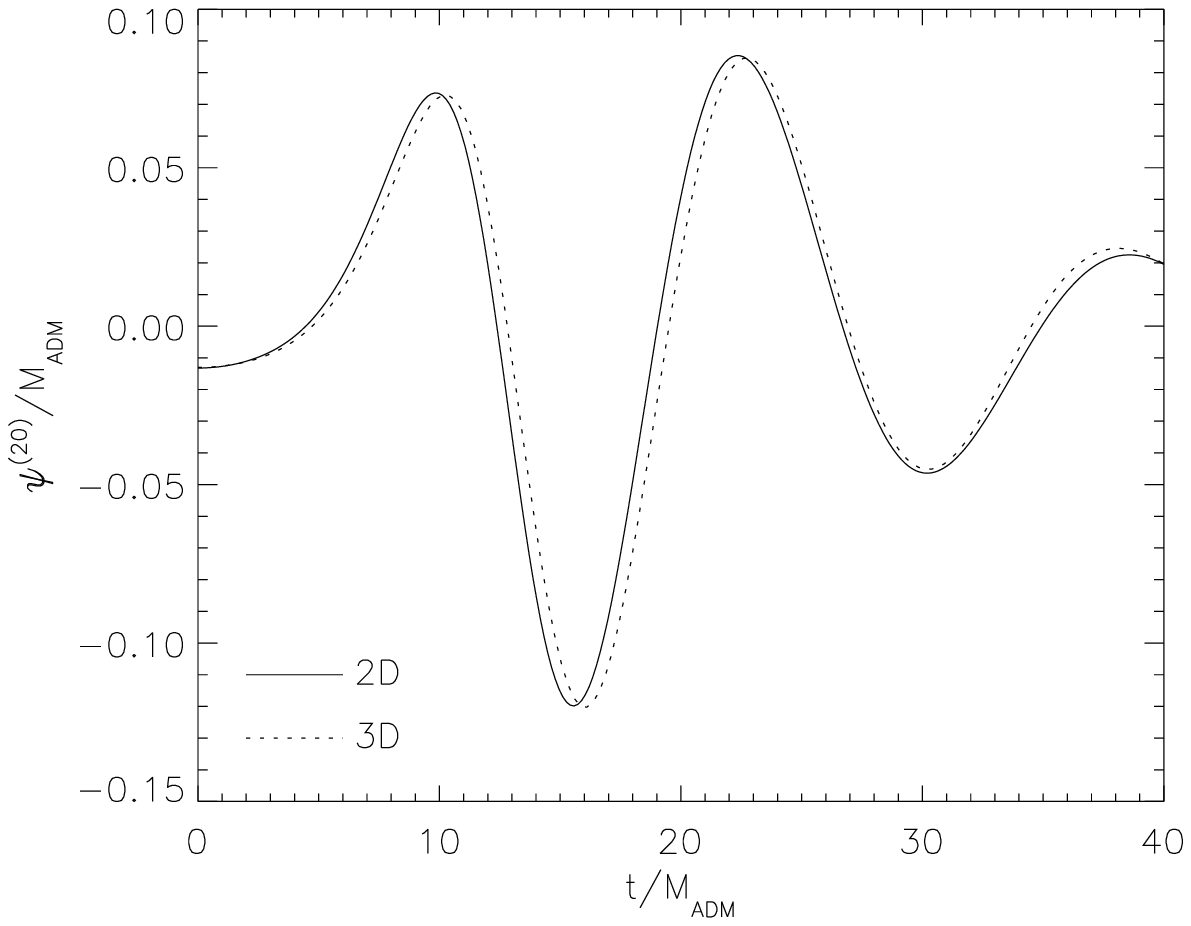}
\epsfxsize=200pt \epsfbox{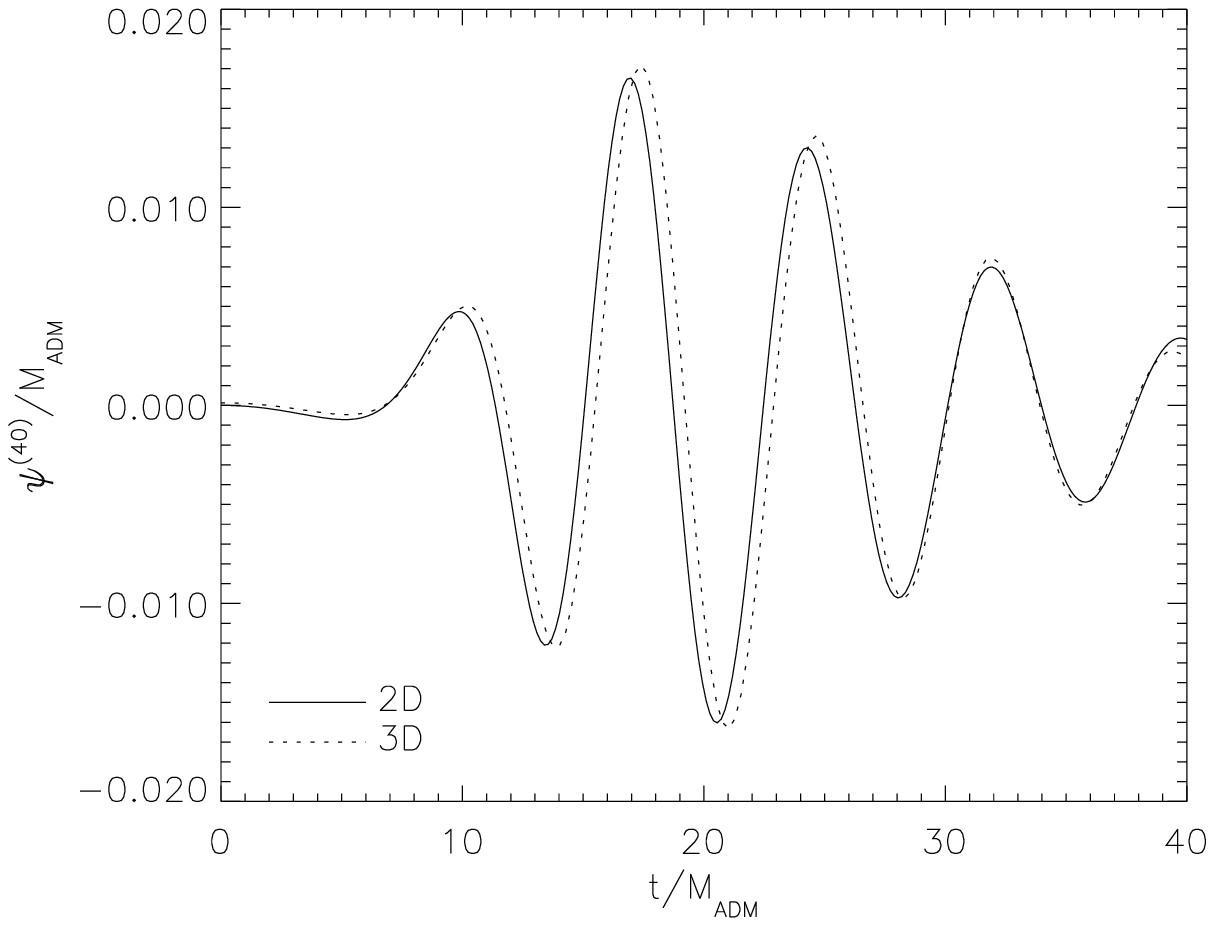}
\epsfxsize=200pt \epsfbox{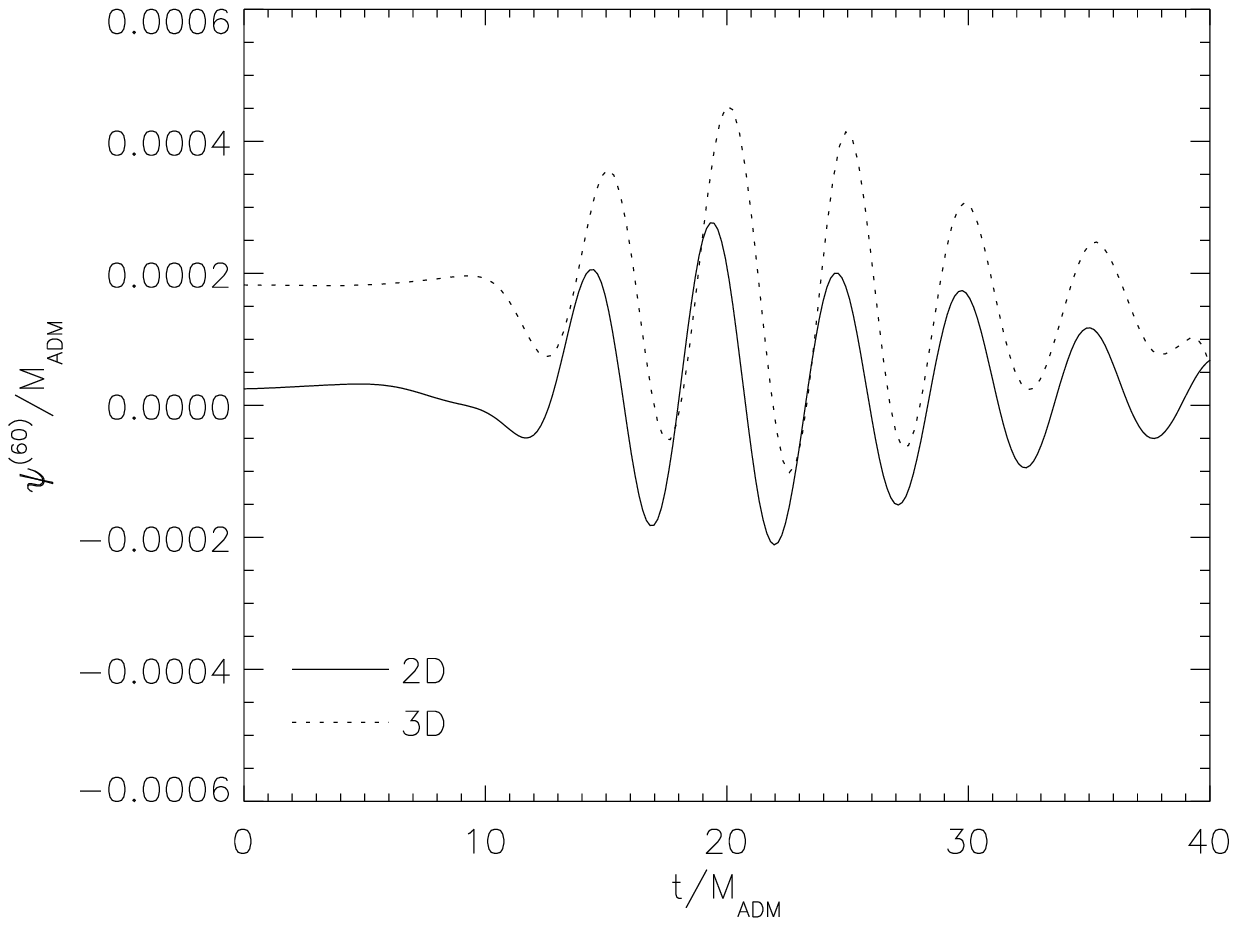}
\caption{We show the (a) $\ell=2$, (b) $\ell=4$, and (c) $\ell=6$
Zerilli functions vs.  time, extracted during 2D and 3D evolutions of
the data set $(a,b,w,n,c)=(0.1,0,1,4,0)$ (Problem 3).  The functions
were extracted at a radius of $7.88M_{ADM}$.  The 2D data were
obtained with $202\times 54$ grid points, giving a resolution of
$\Delta\eta=\Delta\theta=0.03$.  The 3D data were obtained using
$300^3$ grid points and a resolution of $\Delta x=0.104M_{ADM}$.}
\label{fig:run3.zercomp}
\end{figure}

Finally, we study the waveforms extracted in Problem 3, which has a
much lower distortion amplitude, and hence much weaker waves which are
harder to evolve and extract accurately.  In
Fig.~\ref{fig:run3.zercomp} we show the three waveforms.  The $\ell=2$
and $\ell=4$ 3D waveforms agree well with those computed in 2D, as in
previous cases.  But the $\ell=6$ waveform has an offset indicating
some level of error in either the extraction or the evolution itself.
In this simulation the $\ell=6$ mode is about 20 times smaller than
that extracted in Problem 2.  If one examines the $\ell=6$ waveform
from Problem 2 closely, one can see a similar offset of the same
level, but compared to the waveform in that case the effect is almost
not noticeable.  However, in Problem 3, with such a smaller amplitude
signal, the offset can be seed readily.  Resolution studies of this
mode indicate that it is converging toward the 2D result as the
resolution increases.  But even at the highest $300^{3}$ resolution,
there is enough error to make extracting this very small signal (much
smaller than any other modes presented) rather difficult.  The reasons
for the offset in $\ell=6$ waveforms are under investigation.

The energy emitted in the $\ell=2$ mode is $9\times 10^{-5} M_{ADM}$,
again, only a few per cent of the $MRL$. The
energy emitted in the $\ell=4$ mode is $6\times 10^{-6} M_{ADM}$.

\section{Conclusions}
\label{sec:conclusions}

In this followup paper to Ref.~\cite{Camarda97b}, we have performed
the first detailed study of the dynamics of axisymmetric, distorted
black holes with a general 3D code.  This work is another step in the
development of 3D numerical relativity towards codes capable of
simulating the spiraling coalescence of binary black hole systems.  In
this step, we have shown that, given sufficient resolution in 3D
Cartesian coordinates, distorted black holes, which are used to model
the late stages of binary black hole mergers just after the horizons
have merged, can be accurately evolved.  We studied the behavior of
metric functions and apparent horizons in these spacetimes evolved in
3D Cartesian coordinates, and compared them with evolutions provided
by 2D codes in polar-spherical coordinates, and showed that the
results agree well.

Furthermore, we showed that the gravitational waveforms generated by
the black hole, consisting of small perturbations on the evolving
black hole background, can be accurately propagated and extracted from
the numerically generated metric, on a 3D Cartesian grid.  We have
demonstrated this by comparing results from a mature 2D code, showing
good agreement not only for the $\ell=2$, but also with the higher
$\ell=4$ and even the $\ell=6$ modes of the radiation, in cases where
they are strong enough to stand above numerical error.  Such modes are much
weaker than the dominant $\ell=2$ modes, and have much higher
frequency and more complex angular dependence, yet they can be
accurately tracked and resolved in full 3D. To our knowledge, this is
the first study of $\ell=6$ modes with any nonlinear evolution code.

Although we regard this as an important step in establishing numerical
relativity as a viable tool to compute waveforms from black hole
interactions, the calculations one can presently do are limited.  With
present techniques, the evolutions can only be carried out for a
fraction of the time required to simulate the 3D orbiting coalescence.
Many techniques to handle this more general case are under
development, such as hyperbolic formulations of the Einstein equations
and the advanced numerical methods they bring\cite{Bona97a}, adaptive
mesh refinement that will enable placing the outer boundary farther
away while resolving the strong field region where the waves are
generated\cite{Papadapoulos98a}, and apparent horizon boundary
conditions that excise the interiors of the black holes, thus avoiding
the difficulties associated with singularity avoiding
slicings\cite{Seidel92a,Daues96a,Cook97a}.

In a completely different approach, a 3D code based on a
characteristic formulation of the equations was shown to overcome some
limitations of 3D Cauchy codes, to evolve distorted black holes for
essentially unlimited times (more than $t \approx 60,000M$), on
distorted black holes.  Although it is not clear if characteristic
evolution will be able to handle highly distorted or colliding black
holes, it would be very interesting to study black hole systems such
as these to see if the black hole ringdown and waveforms can be
also be accurately extracted.

All of these techniques, and others, may be needed to handle the more
general, long term evolution of coalescing black holes.  Our purpose
in this paper has been to show that {\em (a)} given present resources
one can evolve simpler distorted black hole systems and accurately
extract the waveforms, even when they carry only $10^{-6}M$ in energy,
and {\em (b)} to establish testbeds for the techniques under
development for the more general case.  Experience has shown that
although the gross dynamics of black hole evolutions are fairly robust
to different treatments in fully nonlinear codes, the waveforms are
very sensitive and difficult to compute
correctly~\cite{Camarda97b,Brandt94b,Brandt94c,Abrahams92a,Anninos93c,Anninos93b,Anninos94b,Camarda97a,Anninos94a}.
 Each of these new techniques may introduce numerical artifacts, even
if at very low amplitude, to which the waveforms may be very
sensitive.  As new methods are developed and applied to numerical
black hole simulations, they can now be tested on evolutions such as
those presented here to ensure that the waveforms can be accurately
computed.

In future papers we extend this work and apply it to full 3D initial
data sets where nonaxisymmetric modes can be extracted for the first
time\cite{Camarda97a,Allen98a,Allen98b}, and to the evolution of
colliding black holes in 3D, extending the work in \cite{Anninos96c}.
Once these techniques have been fully developed and tested on true 3D
data sets, it will be important to apply it to true 3D black hole
collision simulations, such as those recently reported by Br{\"u}gmann
\cite{Bruegmann97}.

\acknowledgements This work has been supported by the Albert Einstein
Institute (AEI), NCSA, and the Binary Black Hole Grand Challenge
Alliance, NSF PHY/ASC 9318152 (ARPA supplemented).  We would like to
thank K.V. Rao, John Shalf, and the staff at NCSA for assistance with
the computations.  Among many colleagues at NCSA, AEI, and Washington
University who have influenced this particular work, we especially
thank Andrew Abrahams, Gabrielle Allen, Larry Smarr, and Wai-Mo Suen.
E.S. would like to thank Carles Bona and J. Mass\'o for hospitality at
the University of the Balearic Islands for hospitality while this
manuscript was revised.  Calculations were performed at AEI and NCSA
on an SGI/Cray Origin 2000 supercomputer.


\begin{thebibliography}{10}

\bibitem{Camarda97b}
K. Camarda and E. Seidel, Phys. Rev. D {\bf 57},  R3204  (1998).

\bibitem{Abramovici92}
A.~A. Abramovici {\it et~al.}, Science {\bf 256},  325  (1992).

\bibitem{Flanagan97a}
\'{E}. \'{E}.~Flanagan and S.~A. Hughes, Phys. Rev. D {\bf 57},  4535
  (1998).

\bibitem{Flanagan97b}
\'{E}. \'{E}.~Flanagan and S.~A. Hughes, Phys. Rev. D {\bf 57},  4566
  (1998).

\bibitem{Stark85}
R.~F. Stark and T. Piran, Phys. Rev. Lett. {\bf 55},  891  (1985).

\bibitem{Abrahams94a}
A.~M. Abrahams, G.~B. Cook, S.~L. Shapiro, and S.~A. Teukolsky, Phys. Rev. D
  {\bf 49},  5153  (1994).

\bibitem{Brandt94b}
S. Brandt and E. Seidel, Phys. Rev. D {\bf 52},  856  (1995).

\bibitem{Brandt94c}
S. Brandt and E. Seidel, Phys. Rev. D {\bf 52},  870  (1995).

\bibitem{Abrahams92a}
A. Abrahams {\it et~al.}, Phys. Rev. D {\bf 45},  3544  (1992).

\bibitem{Anninos93c}
P. Anninos {\it et~al.},  in {\em Computational Astrophysics: Gas Dynamics and
  Particle Methods}, edited by W. Benz, J. Barnes, E. Muller, and M. Norman
  (Springer-Verlag, New York, 1997), in press.

\bibitem{Anninos93b}
P. Anninos {\it et~al.}, Phys. Rev. Lett. {\bf 71},  2851  (1993).

\bibitem{Anninos94b}
P. Anninos {\it et~al.}, Phys. Rev. D {\bf 52},  2044  (1995).

\bibitem{Anninos94c}
P. Anninos {\it et~al.}, Phys. Rev. D {\bf 52},  2059  (1995).

\bibitem{Seidel92a}
E. Seidel and W.-M. Suen, Phys. Rev. Lett. {\bf 69},  1845  (1992).

\bibitem{Anninos94e}
P. Anninos {\it et~al.}, Phys. Rev. D {\bf 51},  5562  (1995).

\bibitem{Scheel94}
M.~A. Scheel, S.~L. Shapiro, and S.~A. Teukolsky, Phys. Rev. D {\bf 51},  4208
  (1995).

\bibitem{Marsa96}
R. Marsa and M. Choptuik, Phys Rev D {\bf 54},  4929  (1996).

\bibitem{Daues96a}
G.~E. Daues, Ph.D. thesis, Washington University, St. Louis, Missouri, 1996.

\bibitem{Cook97a}
G.~B. Cook {\it et~al.}, Phys. Rev. Lett. {\bf 80},  2512  (1998).

\bibitem{Gomez98a}
R. Gomez {\it et~al.}, Phys. Rev. Lett. {\bf 80}, 3915 (1998).

\bibitem{Camarda97a}
K. Camarda, Ph.D. thesis, University of Illinois at Urbana-Champaign, Urbana,
  Illinois, 1998.

\bibitem{Allen97a}
G. Allen, K. Camarda, and E. Seidel, in preparation.

\bibitem{Allen98a}
G. Allen, K. Camarda, and E. Seidel, in preparation.

\bibitem{Misner60}
C. Misner, Phys. Rev. {\bf 118},  1110  (1960).

\bibitem{Brill63}
D.~S. Brill and R.~W. Lindquist, Phys. Rev. {\bf 131},  471  (1963).

\bibitem{Bowen80}
J. Bowen and J.~W. York, Phys. Rev. D {\bf 21},  2047  (1980).

\bibitem{Cook90}
G. Cook, Ph.D. thesis, University of North Carolina at Chapel Hill, Chapel
  Hill, North Carolina, 1990.

\bibitem{Cook91}
G.~B. Cook, Phys. Rev. D {\bf 44},  2983  (1991).

\bibitem{Cook93}
G.~B. Cook {\it et~al.}, Phys. Rev. D {\bf 47},  1471  (1993).

\bibitem{Brandt97a}
S. Brandt, K. Camarda, and E. Seidel, in preparation.

\bibitem{Bernstein94a}
D. Bernstein, D. Hobill, E. Seidel, and L. Smarr, Phys. Rev. D {\bf 50},  3760
  (1994).

\bibitem{Bernstein93a}
D. Bernstein, Ph.D. thesis, University of Illinois Urbana-Champaign, 1993.

\bibitem{Brandt94a}
S. Brandt and E. Seidel, Phys. Rev. D {\bf 54},  1403  (1996).

\bibitem{Brill59}
D.~S. Brill, Ann. Phys. {\bf 7},  466  (1959).

\bibitem{Bernstein93b}
D. Bernstein {\it et~al.}, Phys. Rev. D {\bf 50},  5000  (1994).

\bibitem{Anninos93a}
P. Anninos {\it et~al.}, Phys. Rev. D {\bf 50},  3801  (1994).

\bibitem{Anninos93d}
P. Anninos {\it et~al.}, IEEE Computer Graphics and Applications {\bf 13},  12
  (1993).

\bibitem{Anninos94f}
P. Anninos {\it et~al.}, Phys. Rev. Lett. {\bf 74},  630  (1995).

\bibitem{Anninos95c}
P. Anninos {\it et~al.}, Australian Journal of Physics {\bf 48},  1027  (1995).

\bibitem{Anninos96c}
P. Anninos, J. Mass\'o, E. Seidel, and W.-M. Suen, Physics World {\bf 9},  43
  (1996).

\bibitem{Anninos94d}
P. Anninos {\it et~al.}, Phys. Rev. D {\bf 56},  842  (1997).

\bibitem{Anninos96b}
P. Anninos {\it et~al.}, Phys. Rev. D {\bf 54},  6544  (1996).

\bibitem{Balakrishna96a}
J. Balakrishna {\it et~al.}, Class. Quant. Grav. {\bf 13},  L135  (1996).

\bibitem{Libson94b}
P. Anninos {\it et~al.}, gr-qc/9609059,    (1998), to appear in
  Phys. Rev. D.

\bibitem{Zerilli70}
F.~J. Zerilli, Phys. Rev. Lett. {\bf 24},  737  (1970).

\bibitem{Moncrief74}
V. Moncrief, Annals of Physics {\bf 88},  323  (1974).

\bibitem{Abrahams88}
A. Abrahams, Ph.D. thesis, University of Illinois, Urbana, Illinois, 1988.

\bibitem{Gleiser96a}
R.~J. Gleiser, C.~O. Nicasio, R.~H. Price, and J. Pullin, Class. Quant. Grav.
  {\bf 13},  L117  (1996).

\bibitem{Bona97a}
C. Bona, J. Mass\'o, E. Seidel, and J. Stela, Phys. Rev. D {\bf 56},  3405
  (1997).

\bibitem{Papadapoulos98a}
P. Papadopoulos, E. Seidel, and L. Wild, Phys. Rev. D  (1998), submitted,
  gr-qc/9802069.

\bibitem{Anninos94a}
P. Anninos {\it et~al.}, Technical Report No.~24, National Center for
  Supercomputing Applications (unpublished).

\bibitem{Allen98b}
G. Allen, K. Camarda, and E. Seidel, in preparation.

\bibitem{Bruegmann97}
B. Br\"ugmann, gr-qc/9708035.

\end{thebibliography}

\end{document}